\documentclass{aa}
\usepackage{graphicx}
\usepackage{rotating}
\usepackage{longtable}
\begin{document}

\newcommand{\as}[2]{$#1''\,\hspace{-1.7mm}.\hspace{.1mm}#2$}
\newcommand{\am}[2]{$#1'\,\hspace{-1.7mm}.\hspace{.0mm}#2$}
\def\approxlt{\lower.2em\hbox{$\buildrel < \over \sim$}}
\def\approxgt{\lower.2em\hbox{$\buildrel > \over \sim$}}
\newcommand{\dgr}{\mbox{$^\circ$}}   
\newcommand{\grd}[2]{\mbox{#1\fdg #2}}
\newcommand{\gsim}{\stackrel{>}{_{\sim}}}
\newcommand{\HI}{\mbox{H\,{\sc i}}}
\newcommand{\HIbf}{\mbox{H\hspace{0.155 em}{\footnotesize \bf I}}}
\newcommand{\HIit}{\mbox{H\hspace{0.155 em}{\footnotesize \it I}}}
\newcommand{\HIsl}{\mbox{H\hspace{0.155 em}{\footnotesize \sl I}}}
\newcommand{\HII}{\mbox{H\,{\sc ii}}}
\newcommand{\IHI}{\mbox{${I}_{HI}$}}
\newcommand{\Jykms}{\mbox{Jy~km~s$^{-1}$}}
\newcommand{\kms}{\mbox{km\,s$^{-1}$}}
\newcommand{\kmsMpc}{\mbox{ km\,s$^{-1}$\,Mpc$^{-1}$}}
\def\lir{{\hbox {$L_{IR}$}}}
\def\lco{{\hbox {$L_{CO}$}}}
\def \ls{\hbox{$L_{\odot}$}}
\newcommand{\LB}{\mbox{$L_{B}$}}
\newcommand{\LBnul}{\mbox{$L_{B}^0$}}
\newcommand{\LBsun}{\mbox{$L_{\odot,B}$}}
\newcommand{\lsim}{\stackrel{<}{_{\sim}}}
\newcommand{\LsunK}{\mbox{$L_{\odot, K}$}}
\newcommand{\LsunB}{\mbox{$L_{\odot, B}$}}
\newcommand{\LsunMsun}{\mbox{$L_{\odot}$/${M}_{\odot}$}}
\newcommand{\LK}{\mbox{$L_K$}}
\newcommand{\LKLB}{\mbox{$L_K$/$L_B$}}
\newcommand{\LKLBnul}{\mbox{$L_K$/$L_{B}^0$}}
\newcommand{\LKLsun}{\mbox{$L_{K}$/$L_{\odot,Bol}$}}
\newcommand{\masq}{\mbox{mag~arcsec$^{-2}$}}
\newcommand{\MHI}{\mbox{${M}_{HI}$}}
\newcommand{\MHILB}{\mbox{$M_{HI}/L_B$}}
\newcommand{\MHILBfr}{\mbox{$\frac{{M}_{HI}}{L_{B}}$}}
\newcommand{\MHILK}{\mbox{$M_{HI}/L_K$}}
\newcommand{\KMS}{\mbox{$\frac{km}{s}$}}
\newcommand{\JYKMS}{\mbox{$\frac{Jy km}{s}$}}
\newcommand{\MHILKfr}{\mbox{$\frac{{M}_{HI}}{L_{K}}$}}
\def \ms{\hbox{$M_{\odot}$}}
\newcommand{\Msun}{\mbox{${M}_\odot$}}
\newcommand{\MsunLsun}{\mbox{${M}_{\odot}$/$L_{\odot,Bol}$}}
\newcommand{\MsunLBsun}{\mbox{${M}_{\odot}$/$L_{\odot,B}$}}
\newcommand{\MsunLKsun}{\mbox{${M}_{\odot}$/$L_{\odot,K}$}}
\newcommand{\MT}{\mbox{${M}_{ T}$}}
\newcommand{\MTLBnul}{\mbox{${M}_{T}$/$L_{B}^0$}}
\newcommand{\MTLBsun}{\mbox{${M}_{T}$/$L_{\odot,B}$}}
\newcommand{\nan}{Nan\c{c}ay}
\newcommand{\tis}[2]{$#1^{s}\,\hspace{-1.7mm}.\hspace{.1mm}#2$}
\newcommand{\Vcor}{\mbox{$V_{0}$}}
\newcommand{\vhel}{\mbox{$V_{hel}$}}
\newcommand{\VHI}{\mbox{$V_{HI}$}}
\newcommand{\vrot}{\mbox{$v_{rot}$}}
\def\la{\mathrel{\hbox{\rlap{\hbox{\lower4pt\hbox{$\sim$}}}\hbox{$<$}}}}
\def\ga{\mathrel{\hbox{\rlap{\hbox{\lower4pt\hbox{$\sim$}}}\hbox{$>$}}}}

 \title{Completing H{\Large \bf I} observations of galaxies II. The Coma Supercluster}
  
  \author{G. Gavazzi\inst{1},         
          K. O'Neil\inst{2},
	  A. Boselli\inst{3},	  
      \and
	  W. van Driel\inst{4}
          } 

  \offprints{G. Gavazzi}

  \institute{Universit\'a degli Studi di Milano-Bicocca, Piazza delle scienze 3, 
             20126 Milano, Italy \\
            \email{giuseppe.gavazzi@mib.infn.it}
       \and
              NRAO, P.O. Box 2, Green Bank, WV 24944, U.S.A. \\
            \email{koneil@gb.nrao.edu}
      \and   
             Laboratoire d'Astrophysique de Marseille, BP8, Traverse du Siphon, 
             F-13376 Marseille, France \\
            \email{alessandro.boselli@oamp.fr}     	     
       \and
             Observatoire de Paris, Section de Meudon, GEPI, CNRS UMR 8111 
             and Universit\'e Paris 7, 5 place Jules Janssen, F-92195 Meudon Cedex, 
             France \\
            \email{wim.vandriel@obspm.fr}
             }


  \abstract{ {\rm
High sensitivity 21-cm \HI\ line observations, 
with an rms noise level of $\sim 0.5$ mJy, were made of
35 spiral galaxies in the Coma Supercluster, using the refurbished Arecibo telescope, 
which resulted in the detection of 25 objects.
These data, combined with the measurements available from the literature, provide the set 
of \HI\ data for 94 \% of all late-type galaxies in the Coma Supercluster with an apparent photographic magnitude
$m_p \leq 15.7$ mag.
We confirm that the typical scale of \HI\
deficiency around the Coma cluster is 2 Mpc, i.e. one virial radius.
Comparing the \HI\ mass function (HIMF) of cluster with non-cluster members of the Coma Supercluster
we detect a shortage of high \HI\ mass galaxies among cluster members that
can be ascribed to the pattern of \HI\ deficiency found in rich clusters.    
}
   }

 \authorrunning{Gavazzi et al.}
 \titlerunning{\HI\ observations in the Coma Supercluster}
 
 \maketitle

\begin{figure*}
\centering
\includegraphics[width=19.0cm]{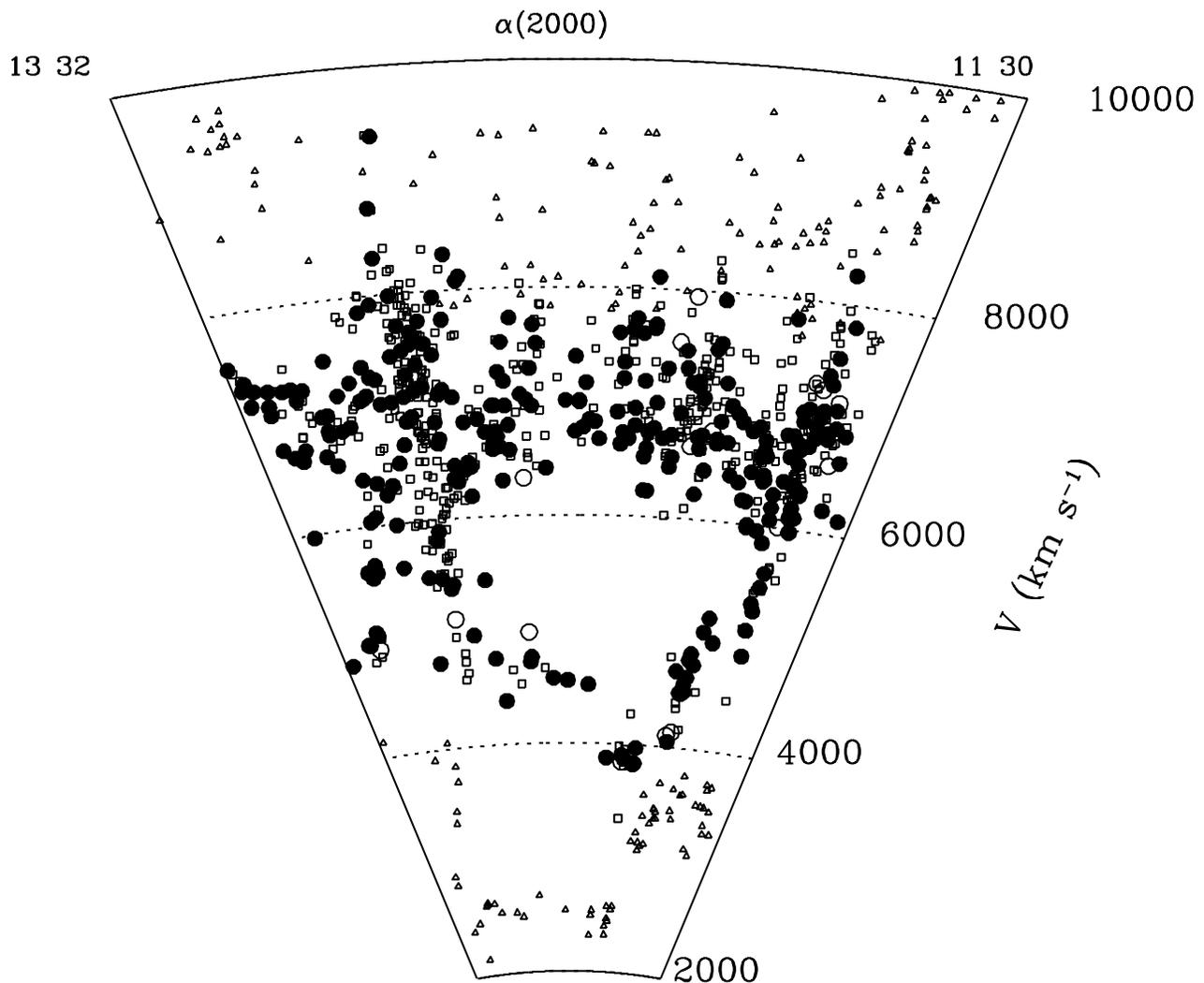}
\caption{The wedge diagram of the Coma Supercluster galaxies in the declination interval $18^{\circ}<\delta<32^{\circ}$.
Members are divided among late-type (large circles, filled if observed in HI) and early-type (E-S0a: empty squares).
Empty triangles mark foreground and background galaxies.}
\label{wedge}
\end{figure*}
\section{Introduction}  
\HI\ line observations of galaxies have provided us with some of the most 
powerful diagnostics on the role of 
the environment in regulating the evolution of late-type (spiral) galaxies in the local Universe. 
Spiral galaxies in rich  X-ray luminous clusters display a significant lack of \HI\ gas  
with respect to their ``undisturbed'' counterparts in the field (Haynes \& Giovanelli 1984, Giovanelli \&  Haynes 1985).
This pattern of \HI\ deficiency can be attributed 
to various interaction mechanisms: ram-pressure (Gunn \& Gott 1972),
viscous stripping (Nulsen 1982), thermal evaporation (Cowie \& Songaila 1977) or tidal 
interaction with the cluster potential well (Byrd \& Valtonen 1990;
Moore et al. 1996)). Since these mechanisms have a higher efficiency in or near rich cluster cores, 
the \HI\ deficiency 
parameter (Haynes \& Giovanelli 1984 - see Sect. 5.1) is an environmental indicator that provides a clear signature of
a galaxy's membership of a rich cluster.\\
The Coma Supercluster, due to its proximity to us ($\sim$ 100 Mpc), 
has received considerable attention in \HI\ studies. Since the pioneering study by Sullivan et al. (1981), 
various works (e.g. Chincarini et al. 1983a; Gavazzi 1987; Gavazzi 1989: Scodeggio \&  Gavazzi 1993;
Haynes et al. 1997) have provided measurements of the \HI\ content for most late-type galaxies
in the Coma Supercluster. 
In addition to these single-dish studies of their global \HI\ 
properties, the detailed mapping of galaxies in the Coma and A1367 clusters with radio synthesis telescopes was obtained by
Bravo-Alfaro et al., (2000, 2001) and Dickey \& Gavazzi, (1991).\\ 
A high sensitivity,  blind \HI\ survey of 7000 square deg. of sky is planned for 2005-2006 with the ALFA
multibeam system at Arecibo (Giovanelli et al. 2005), 
and even more sensitive (1 mJy rms) surveys of parts of these clusters will be obtained with the ALFA system. 
They will include the Coma Supercluster and the Virgo cluster.
In preparation for these surveys, before the installation of the ALFA system, we used the single-beam Arecibo system for the continuation of 
the pointed observation survey of late-type galaxies in the Virgo cluster (see Gavazzi et al. 2005a, Paper I) 
and in the Coma Supercluster area.
Here we report on the results of the Coma Supercluster observations which, in conjunction with the previously available
\HI\ data-set, enable us to review the properties of galaxies in this Supercluster, as obtained
from optically selected \HI\ observations. \\
The selection of the cluster targets for \HI\ observations
is described in Section 2, the observations and the data reduction are presented 
in Section 3 and the results are given in Section 4 and discussed in Section 5.
A Hubble constant of $\rm 75 ~km ~s^{-1} ~Mpc^{-1}$ is assumed throughout this paper.

\section{Sample selection} 

Galaxies in the present study are selected from the CGCG Catalogue (Zwicky et al. 1961-68) in the region
$11^h30^m<\alpha<13^h30^m; 18^{\circ}<\delta<32^{\circ}$.
There are 1127 CGCG galaxies listed in this region with an apparent photographic magnitude $m_p\le$15.7.
Their wedge-diagram is given in Fig. 1 where the structure of the Coma--A1367 Supercluster stands out clearly  
as the pronounced density enhancement near 7000 \kms, as part of one of the largest known coherent 
structures  in the local universe, named the "Great Wall" (Ramella et al. 1992).
Other conspicuous features are: the "Fingers of God" of 
Coma and A1367, spanning the interval $4000<V<10000$ \kms, mostly traced by early-type objects,
and the large ($\rm \sim 7500 ~Mpc^3$) "void" in front of the Supercluster, with a density 150 times lower than
the mean galaxy density in the universe. 
Other remarkable features, also known as the "legs of the homunculus" (de Lapparent et al. 1986) 
are two filaments pointing toward the Coma cluster in the interval $4500<V<7000$ \kms.
A third "homunculus leg" surrounds the void on the western side, projected near A1367.  
Objects in the interval $6000<V<8000$ \kms~ form a bridge between the two clusters with  
a narrow velocity distribution. 
Above  $V>8000$ \kms~ the Coma Supercluster fades into the background, and 
setting a boundary between its members and the projected background objects
is rather arbitrary.
In assigning the membership of individual galaxies to the various sub-structures 
we follow the criteria of Gavazzi et al. (1999).\\
Out of the 1127 galaxies in Fig. 1, 654 are considered proper Supercluster members according to these
criteria, and 76 additional galaxies belonging to the "homunculus legs" (HL) are considered separately (see Table 1). 
Their sky projection is given in Fig. 2, revealing a substantial morphology segregation 
between late-type (spiral) and early-type (E-S0a) objects.\\ 
\begin{figure*}
\centering
\includegraphics[width=19.0cm]{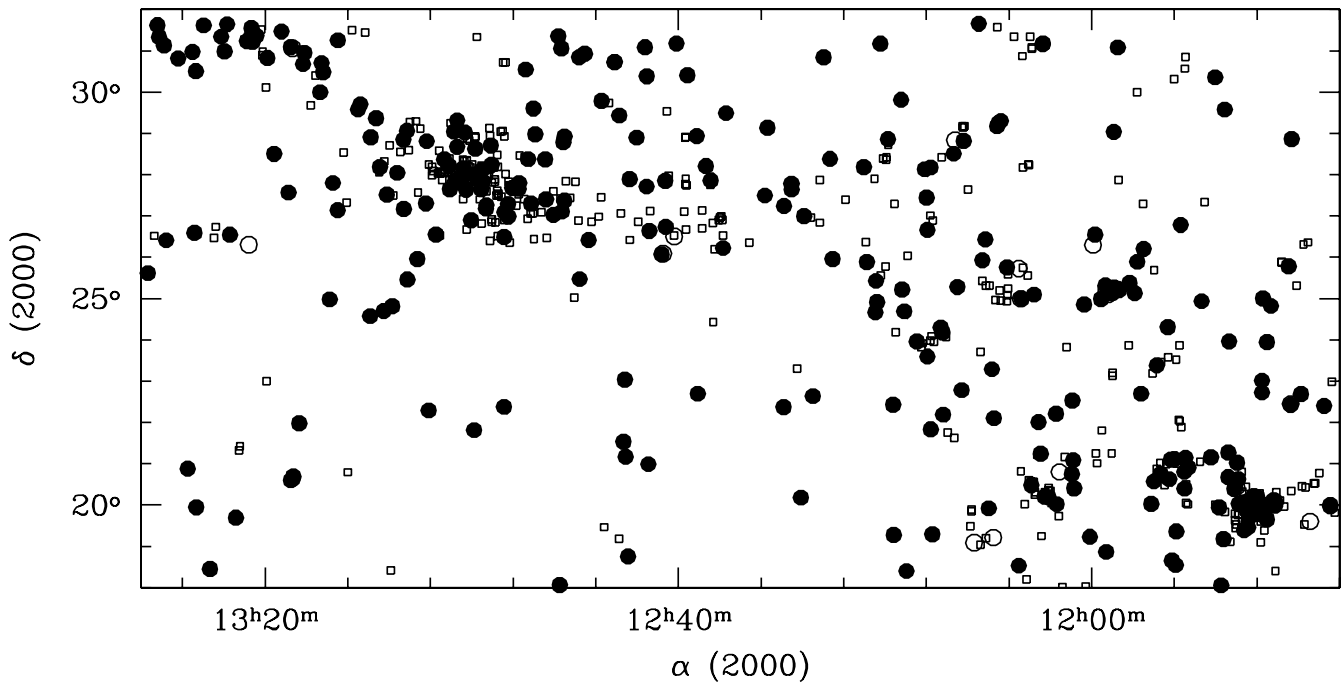}
\caption{The sky projection of Coma Supercluster members (including the "homunculus legs" HL). 
Spirals are represented by circles (filled if observed in \HI) and E-S0a objects by empty squares.}
\label{radec}
\end{figure*}
The Coma Supercluster has been observed in \HI~ with remarkable completeness: 
65 \% of the published data
are found in 6 publications: Chincarini et al. (1983a);
Gavazzi (1987, 1989), Scodeggio \& Gavazzi (1993); Haynes et al. (1997).\\
After the present high sensitivity (rms $\sim 0.5$ mJy) observations of 33 additional CGCG galaxies (and of two
fainter objects: FOCA 610, 636) 295/315 (94 \%) late-type members (including HL) are observed (see Table 1), of which 259 
were detected and 36 have upper limits. 
Of the remaining 20 unobserved late-type targets 13 are members of double or multiple systems 
that could not be resolved
by the Arecibo beam (97-111S, 97-129E, 127-051N, 128-029E, 127-025N, 159-049S, 128-031N,
128-002, 97-036, 98-072, 98-073, 98-081, 98-087). The  
Coma cluster (160-243) and 6 galaxies (127-121, 129-016, 157-077, 158-046, 160-180, 161-029) belonging to the HL
were not observed due to scheduling constraints.
To fill in a scheduling hole, 13 galaxies in the Virgo cluster were also observed.  
All but one were detected. The 
results of these observations are given in Table 3 and the HI profiles of the 
detected galaxies are shown in Figure 8.  These objects will not be  
considered further in this paper.\\
All data on the Coma Supercluster and Virgo galaxies are collected and made available worldwide 
via the "Goldmine" website (http://Goldmine.mib.infn.it; see Gavazzi et al. 2003). 
\begin{table}
{\footnotesize
\begin{tabular}{lllll}
\multicolumn{5}{l}{\footnotesize {\bf Table 1} Sample completeness.} \\
\smallskip \\
\hline
\vspace{-2 mm} \\
  & \multicolumn{2}{c}{Type$<$Sa} & \multicolumn{2}{c}{Type$\geq$Sa}\\ 
  & All & \HI\ & All & \HI\ \\   
\hline
\vspace{-2 mm} \\
Coma members    & 391 & 58 & 263 & 249 \\   
HL         & 24  & 10  & 52 &  46 \\
Background & 131 & 5  & 118 &  55 \\    
Foreground & 46  & 16 & 102 & 92 \\   
All        & 592 & 89 & 535 & 442 \\   
\vspace{-2 mm} \\                        		        				      
\vspace{-2 mm} \\
\hline
\end{tabular}
}
\end{table}

\section{Observations} 

Using the refurbished 305-m Arecibo Gregorian radio telescope we observed 35 galaxies
in the Coma Supercluster (plus 13 in the Virgo cluster)  (see Section 2) in February 2004 and January-March 2005. 
Data were taken with the L-Band Wide receiver, using nine-level sampling with two of 
the 2048 lag subcorrelators set to each polarization channel. All observations were taken 
using the position-switching technique, with each blank sky (or OFF) position observed for 
the same duration, and over the same portion of the telescope dish as the 
on-source (ON) observation. Each 5min+5min ON+OFF pair was followed by a 10s ON+OFF 
observation of a well-calibrated noise diode. The overlaps between both sub-correlators 
with the same polarization allowed a contiguous radial velocity search range of sufficient width from
 -1000 to 8500 \kms. 
The velocity resolution was 2.6 \kms, the instrument's HPBW at 21 cm is 
\am{3}{5}$\times$\am{3}{1} 
and the pointing accuracy is about 15$''$. The pointing positions used
are the optical center positions of the target galaxies listed in Table 1.
Flux density calibration corrections are good to within 10\% (and often much better), see the 
discussion of the errors involved in O'Neil (2004).\\  
Using standard IDL data reduction software available at Arecibo, corrections were applied 
for the variations in the gain and system temperature with zenith angle and azimuth. A 
baseline of order one to three was fitted to the data, excluding those velocity ranges 
with \HI\ line emission or radio frequency interference (RFI). The velocities were corrected 
to the heliocentric system, using the optical convention, and the polarizations were 
averaged. All data were boxcar smoothed to a velocity resolution of 12.9 \kms\ for further 
analysis. For all spectra the rms noise level was determined and for the detected objects 
the central line velocity, the line widths at, respectively, the 50\% and 20\% level of the peak, 
and the integrated line flux were determined.
No flux correction depending on the source size was applied because the optical extent of all detected targets does not
significantly exceed the Arecibo HPBW.

\section{Results} 

\begin{figure*}
\centering
\includegraphics[width=19.0cm]{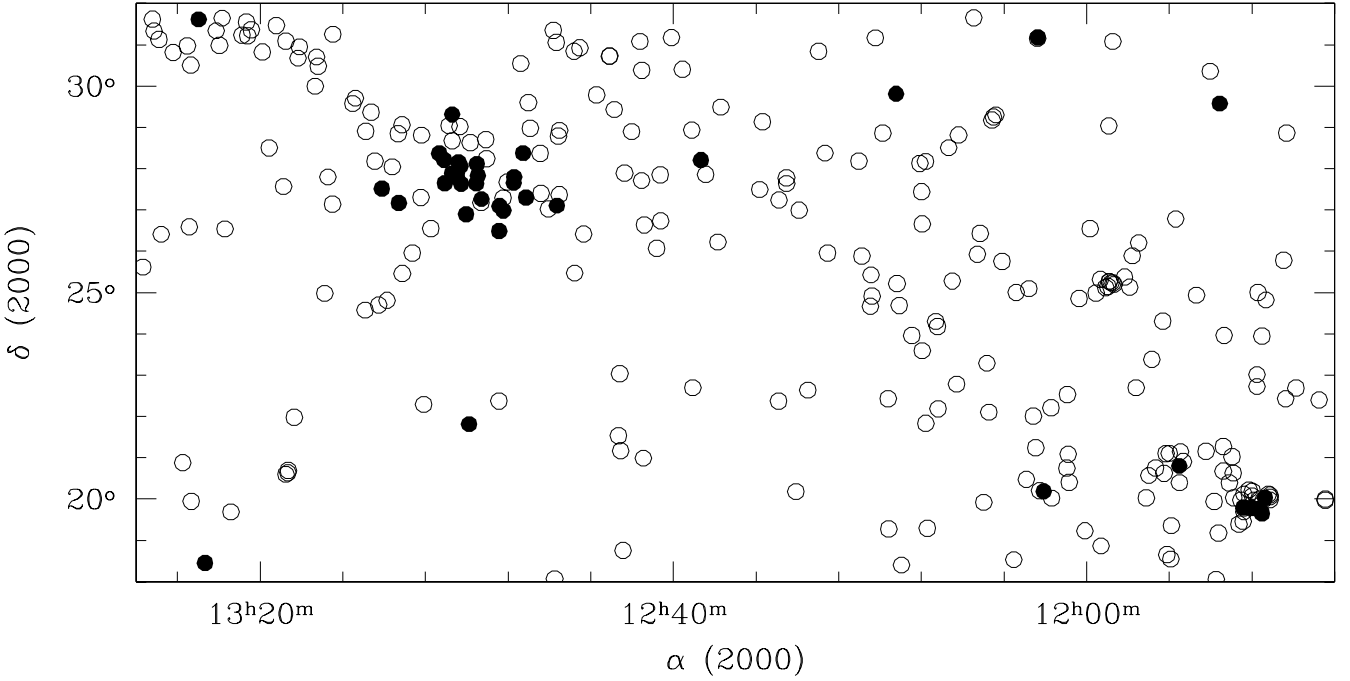}
\caption{The late-type Coma Supercluster members (including HL) coded according to their \HI\ deficiency parameter:
$Def_{HI} \leq 0.5$ (empty circles); $Def_{HI} > 0.5$ (filled circles).}
\label{def}
\end{figure*}

In order to identify sources whose \HI\ detections could have been confused by nearby galaxies,
we queried the NED, HyperLeda and Goldmine databases and inspected DSS images over a region of 10$'$ 
radius surrounding the central position of each source, given the telescope's sidelobe pattern. 
Quoted values are weighted averages from the HyperLeda database, unless otherwise indicated.\\
The \HI\ spectra of both the clearly and the marginally detected galaxies are shown in 
Figures \ref{spectracoma},\ref{spectravcc} 
and the global \HI\ line parameters are listed in Table 3. 
These are directly measured values; 
no corrections have been applied to them for, e.g., instrumental resolution.
Table 3 is organized as follows:\\
Column 1: Obj. is the galaxy designation; \\ 
Column 2-3: (J2000) celestial coordinates;\\
Column 4: the heliocentric optical recessional velocity (in \kms);\\ 
Column 5: the rms dispersion in the baseline {\bf (mJy/beam)}; \\ 
Column 6: $S_p$ is the peak flux density of the detected line {\bf (mJy/beam)}; \\ 
Column 7: $V_{HI}$ is the heliocentric central radial velocity of a line profile 
(in \kms), in the optical convention, with its estimated uncertainty (see below); \\
Columns 8-9: $W_{50}$ and $W_{20}$ are the line widths at 50\% and 20\% of peak maximum, respectively, (\kms); \\ 
Column 10: $I_{HI}$ is the integrated line flux (\Jykms), with its estimated uncertainty (see below).\\  
Column 11: A quality flag to the spectra is given, where
Q=1 stands for high signal-to-noise, double horned profiles, 
Q=2 for high signal-to-noise, single horned profiles, and 
Q=3,4 for low signal-to-noise profiles whose measured line parameters are not reliable.
Q=5 is given to unpublished profiles. \\ 
We estimated the uncertainties 
$\sigma_{V_{HI}}$ (\kms) in \VHI\ and
$\sigma_{I_{HI}}$ (\Jykms) in \IHI\ following Schneider et al. (1986, 1990), as:
\begin{equation}
\sigma_{V_{HI}} = 1.5(W_{20}-W_{50})X^{-1}
\end{equation}
and 
\begin{equation}
\sigma_{I_{HI}} = 2(1.2W_{20}/R)^{0.5}R\sigma = 7.9(W_{20})^{0.5}\sigma 
\end{equation}
where \IHI\ is the integrated line flux (Jy~\kms), 
$R$ is the instrumental resolution (12.9 \kms), and
$X$ is the signal-to-noise ratio of a spectrum, i.e. the ratio of the 
peak flux density $S_p$ and $\sigma$, the rms dispersion in the baseline (Jy). \\
{\bf The uncertainties in the $W_{20}$ and $W_{50}$ line widths are expected to be
2 and 3 times $\sigma_{V_{HI}}$, respectively}.\\

\section{Discussion}

The newly obtained \HI\ data were combined with those available from the
literature for the $m_p \leq 15.7$ late-type galaxies in the Coma Supercluster, as listed in Table 4.
The sample comprises 315 galaxies, of which 295 were observed, 259 were detected and 36 remain undetected.
Table 4 is organized as follows:\\
Column 1: Galaxy designation in the CGCG Catalog;\\
Column 2: Morphological type;\\
Column 3: Apparent photographic magnitude from the CGCG;\\
Column 4: membership
as defined in Gavazzi et al. (1999). Distances $D$ of 
96.0 and 91.3 Mpc are assumed for Coma and A1367 respectively. 
Distances of individual groups and substructures in the Great Wall (including HL) are taken from
Gavazzi et al. (1999) (rescaled to $\rm Ho=75 ~km ~s^{-1} ~Mpc^{-1}$).
Distances from individual redshifts are assumed for supercluster isolated galaxies and members of multiplets.\\
Column 5: recessional velocity, in \kms;\\
Column 6: \HI\ mass or mass limit in solar units: \MHI = 2.36 $10^5 D^2$ \IHI.
For undetected galaxies we set \IHI=$1.5 \times rms_{HI} \times W_{<20-50>}$,
where $rms_{HI}$ is the rms of the spectra in mJy and the $W_{<20-50>}$ profile width is based on the 
following average line widths of the detected objects per Hubble type bin: 
300 \kms\ for Sa-Sbc, 190 \kms\ for Sc-Scd;\\
Column 7: Coded reference to the \HI\ measurement (see the notes to the Table);\\
Column 8: \HI\ deficiency parameter as defined in Haynes \& Giovanelli (1984) (see Sect. 5.1);\\
Column 9: Quality flag (see last Column of Table 3).\\

\subsection{The pattern of \HI\ deficiency}

\begin{figure}
\centering
\includegraphics[width=9.0cm]{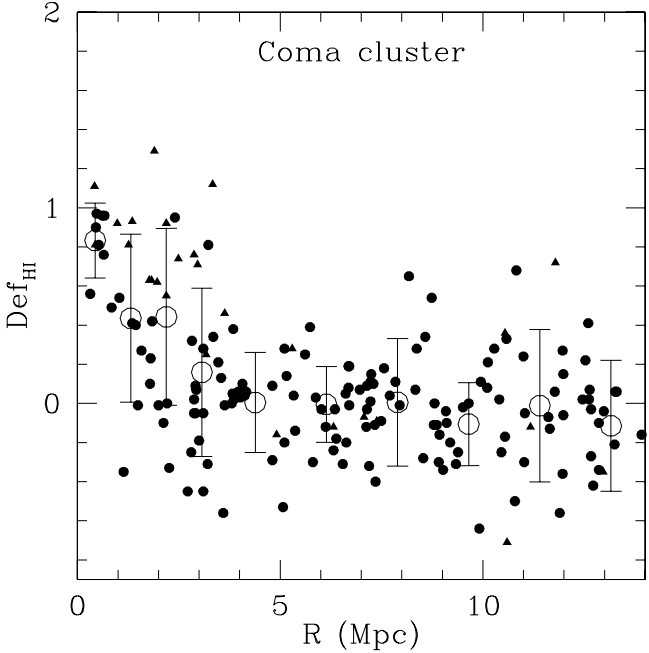}
\caption{The projected distribution about the X-ray center of the Coma cluster of the \HI\ deficiency 
of Late-type Supercluster detected (dots) and undetectd (triangles) members (excluding HL). 
Large circles represent averages in bins of 0.5\dgr~ (1\dgr)  One $\sigma$ errorbars are given.}
\label{comadef}
\end{figure}

For the late-type galaxies in the present study we estimate the \HI\ deficiency parameter 
following Haynes \& Giovanelli (1984) 
as the logarithmic difference between $M_{HI}$ of a reference sample of isolated 
galaxies and $M_{HI}$ actually observed in individual objects: $Def_{HI}= Log M_{HI~ref.} - Log M_{HI~obs.}$. 
$M_{HI~ref}$ is computed from the galaxies optical linear diameter $d$ 
as: $Log M_{HI~ref}=a+b Log(d)$,
where $d$ is estimated consistently with Haynes \& Giovanelli (1984),
$a$ and $b$ are weak functions of the Hubble type, as listed in Table 3 of Paper I
(notice that $b \sim 2$ across the Hubble sequence, i.e. $M_{HI~ref}$ increases approximately 
as the galaxy linear diameter squared).\\ 
Fig.\ref{def} shows that \HI\ deficient galaxies segregate around the center of the Coma cluster, and to a much lesser
degree around A1367. One important issue that can be addressed with
the present data-set, given its high completeness, is on what scale the phenomenon of \HI\ ablation holds.
The \HI\ deficiency parameter of individual galaxies is given in Fig.\ref{comadef} as a function of the  projected
linear separation from the X-ray center of the Coma cluster, out to 15 Mpc, along with average values
taken in bins of 0.5\dgr (from 0\dgr to 2\dgr) and in bins of 1\dgr~ further out (see also Table 2).  It is apparent  
that significant \HI\ deficiency occurs out to approximately 3 Mpc radius. At one virial radius 
(i.e. at 2.2 Mpc, Girardi et al. 1998; or 2.9 Mpc Lokas \& Mamon 2003; Neumann et al. 2001), 
the average \HI\ content of the supercluster galaxies becames indistinguishable from that of 
the field, in agreement with Solanes et al. (2001)\footnote {Galaxies in the HL, (being identified as such outward of 2\dgr~ from Coma and A1367), on average do 
not show significant deficiency (they are not included in Fig.\ref{comadef})}.\\
The star formation rate, as derived from
H$\alpha$ observations, show a significant cut-off near 1-2 projected virial radii 
from the Coma cluster (Gavazzi et al. 2005b) and
from over-density peaks in the SDSS (Nichol 2004; Miller 2004).
What the \HI\ observations show is the driver of the SFR decline, i.e. the lack of gas to sustain it. 

\begin{table*}
{\footnotesize
\begin{tabular}{ccc}
\multicolumn{3}{l}{\footnotesize {\bf Table 2} The radial distribution
of the mean binned \HI\ deficiency 
of Late-type Supercluster members (excluding HL) about the Coma cluster.} \\
\smallskip \\
\hline
\vspace{-2 mm} \\
   $\Theta$(Deg.) & R (Mpc) & $<Def_{HI}>$  \\   
\hline
\vspace{-2 mm} \\
0.0-0.5 &     0.42   & 0.83$\pm$ 0.19\\
0.5-1.0 &     1.25   & 0.43$\pm$ 0.43\\
1.0-1.5 &     2.09   & 0.44$\pm$ 0.45\\
1.5-2.0 &     2.93   & 0.16$\pm$ 0.43\\
2.0-3.0 &	4.19   & 0.00$\pm$ 0.25\\
3.0-4.0 &	5.87   & 0.00$\pm$ 0.19\\
4.0-5.0 &	7.55   & 0.00$\pm$ 0.32\\
5.0-6.0 &	9.24   &-0.10$\pm$ 0.21\\
6.0-7.0 &	10.9   &-0.01$\pm$ 0.39\\
7.0-8.0 &	12.6   &-0.11$\pm$ 0.33\\
\vspace{-2 mm} \\                        		        				      
\vspace{-2 mm} \\
\hline
\label{angdef}
\end{tabular}
}
\end{table*}

\subsection{The \HI\ mass--$B_T^o$ relation}

\begin{figure}
\centering
\includegraphics[width=9.0cm]{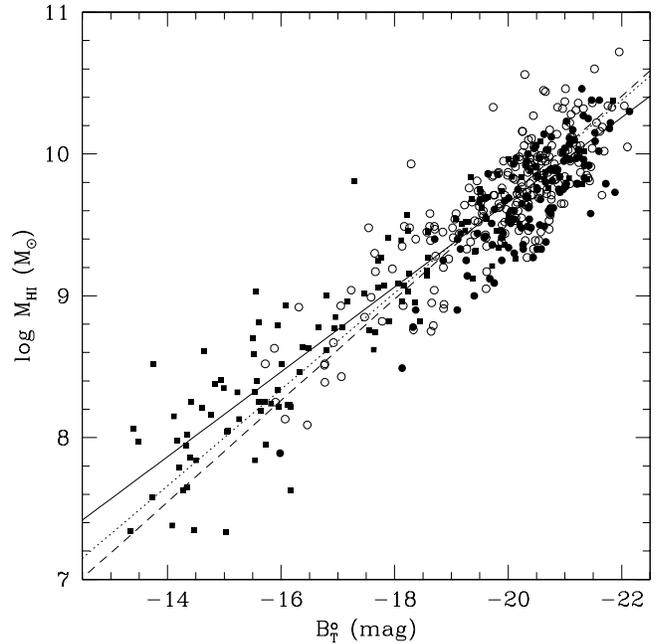}
\caption{The correlation between log $M_{HI}$ and $B_T^o$ for Sa-Sb (filled circles), Sbc-Sc (empty circles),
Scd-Im-BCD (filled squares). The direct fit (solid line),
the inverse fit (dashed) (obtained considering all Hubble types) and the relation used in Paper I (dotted) are given.}
\label{MBMHI}
\end{figure}

Taking advantage from the large sample of optically selected galaxies
with \HI\ measurements and  optical (B-band) photometry (accurate to within 0.1 mag)
we study the \HI\ mass vs. $B_T^o$ relation for unperturbed galaxies.
However since the Coma Supercluster sample ($m_p\leq 15.7$) contains only giant galaxies
brighter than $M_p= -19.1$, for the purpose of extending this correlation over a broader magnitude range
we have included the Virgo galaxies (Paper I). In doing so we must obviously exclude galaxies
with perturbed \HI\ contents. Conservatively we exclude galaxies with
$Def_{HI}\geq 0.2$
(we recall that the $Def_{HI}$ parameter is determined from diameters (see Section 5.1), 
independently of the B luminosity, as in Haynes \& Giovanelli 1984). 
Fig. \ref{MBMHI} shows the relation for 465 late-type galaxies which were selected accordingly,
plotted in three bins of Hubble type to stress the consistency among them. 
The three given linear regressions, obtained combining all Hubble types,  
are the direct one ($M_{HI}=3.680-0.299 \times M_p$), the inverse one ($M_p=7.090-2.795 \times M_{HI}$)
and the one adopted in Paper I ($M_{HI}=2.9-0.34 \times M_p$).
The residual of the (direct) correlation is $\sigma(logM_{HI})$=0.26, i.e., the
\HI\  mass of disk galaxies can be predicted within a factor of 1.8 uncertainty from their B luminosity.

\subsection{The \HI\ mass function}

\begin{figure}
\centering
\includegraphics[width=9.0cm]{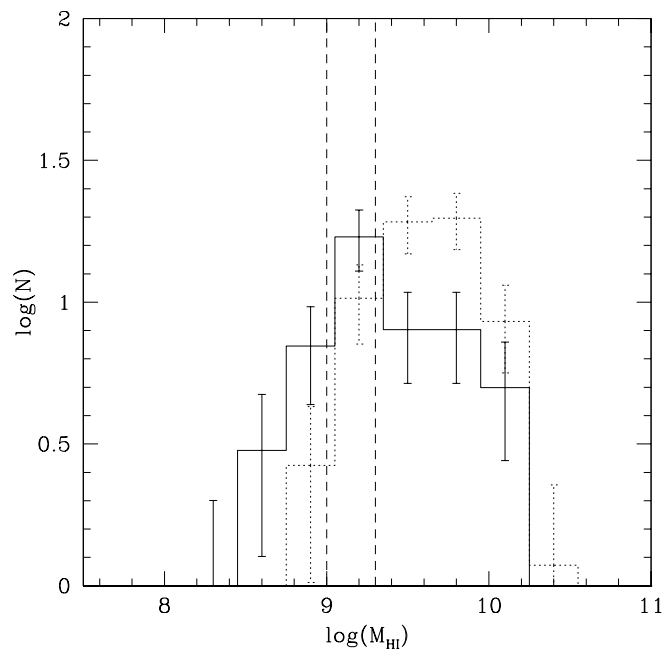}
\caption{The MHI function for the members of the Coma+A1367 clusters (solid) and for the non-cluster
Supercluster members (dotted).
The optical (log$M_{HI}$=9.3 $M_\odot$) and radio (log$M_{HI}$=9.0 $M_\odot$) completeness limits are drawn. }
\label{MHIF}
\end{figure}

\begin{figure*}
\centering
\includegraphics[width=19.0cm]{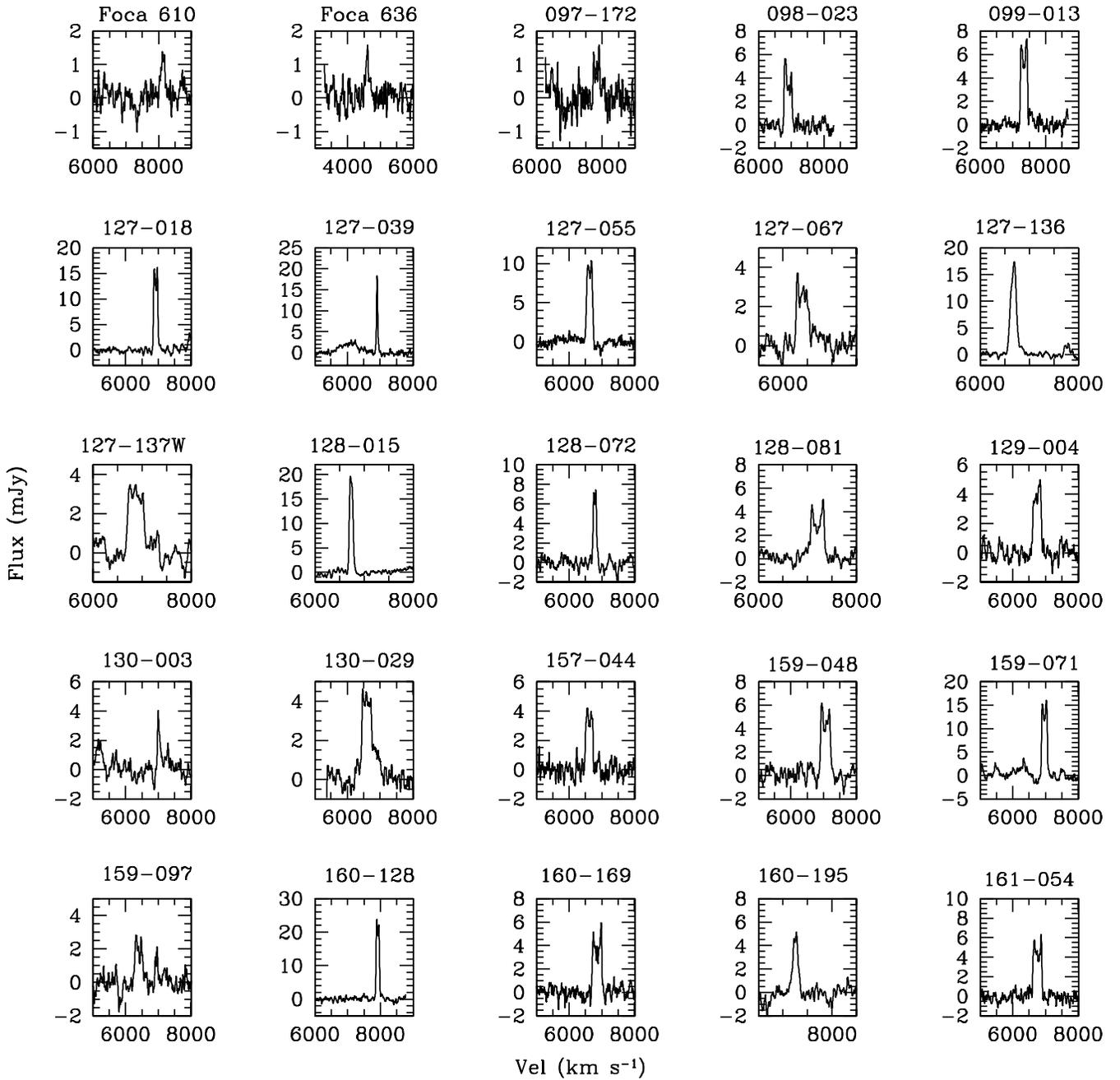}
\caption{\HI\ spectra of the tentatively detected galaxies in the Coma Supercluster.}
\label{spectracoma}
\end{figure*}

\begin{figure*}
\centering
\includegraphics[width=19.0cm]{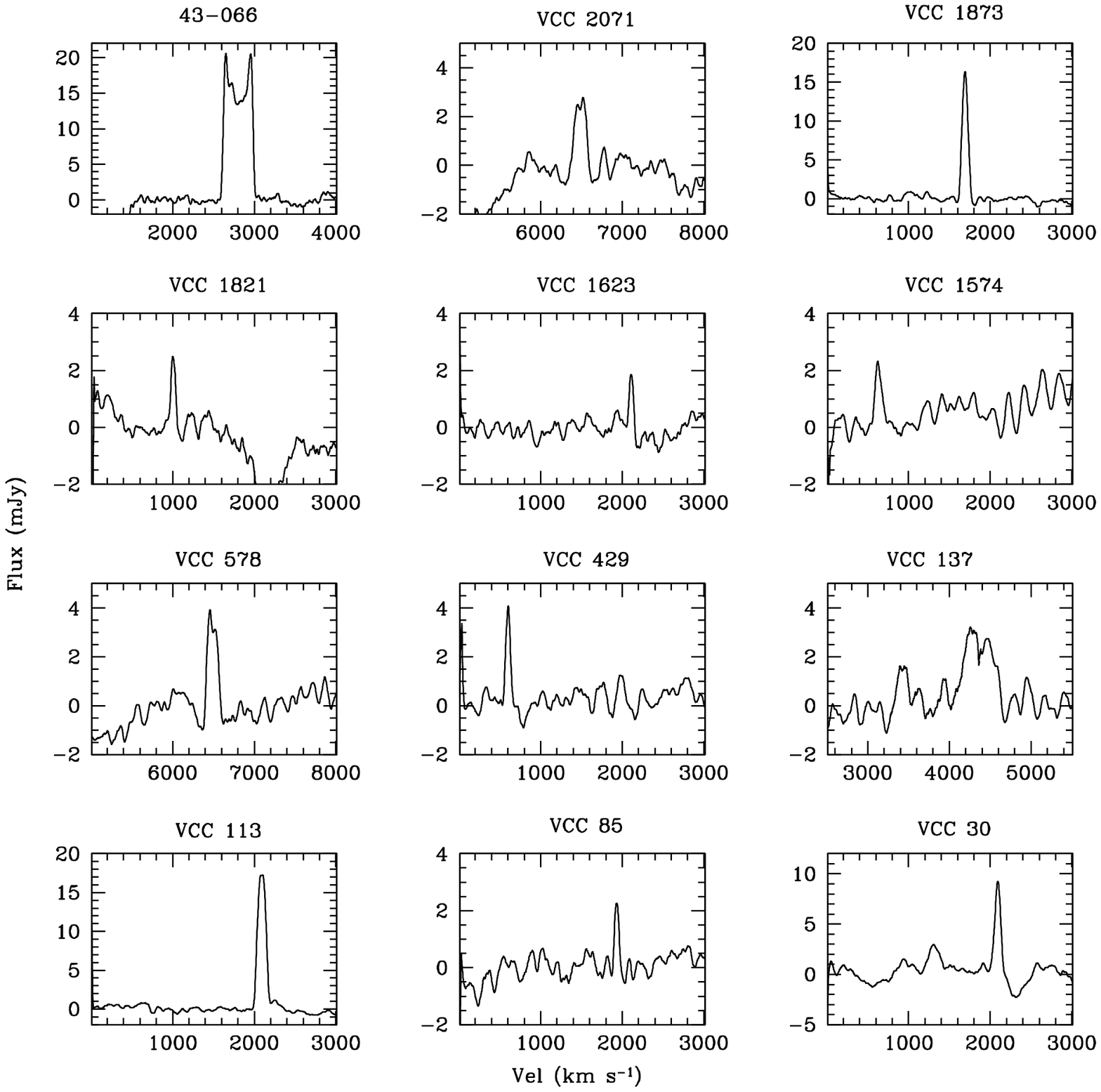}
\caption{\HI\ spectra of the tentatively detected galaxies in the Virgo cluster.}
\label{spectravcc}
\end{figure*}

Fig. \ref {MHIF} shows the frequency distribution of $M_{HI}$ for the cluster (Coma + A1367) late-type 
galaxies (solid histogram) and for
the isolated galaxies (including HL) (the latter normalized to the former by the ratio of the number
of galaxies). The \HI\ Mass Function (HIMF) so obtained cannot be meaningfully compared with the one of Virgo, 
obtained in Paper I,
nor with the one of isolated galaxies by Zwaan et al. (2003) because 
of the shallowness of the optical and the 21cm observations available for the Coma Supercluster.
Given the relation $M_{HI}=3.680-0.299 \times M_p$, discussed in the previous section, 
the limiting magnitude $m_p\leq 15.7$ ($M_p\geq -19.1$) of the CGCG
implies that on average only galaxies with log$M_{HI}>9.3M_\odot$ are targeted. 
This imposes a completeness cut-off that is even shallower than the limiting
detectable \HI\ mass of log$M_{HI}>$9.0 $M_\odot$ that derives from the 
typical noise figure of the 21cm observations ($<\sigma> \sim 1$ mJy).
In Paper I we showed that significant differences between the HIMF of the Virgo cluster and of the field
occur for log$M_{HI}<$9.0 $M_\odot$, i.e. below both the present cut-off lines.
For $9.5 < log M_{HI} < 10$ $M_\odot$ there is however in Fig. \ref {MHIF} a barely 
significant excess in the frequency of non-cluster
members of the Coma Supercluster with respect to cluster members. 
This can be understood as a signature of the deficiency pattern of cluster galaxies, as found in Paper I for Virgo.

\section{Summary and conclusions}

We have observed in the 21-cm \HI\ line, with the refurbished Arecibo telescope, 35 galaxies in the Coma Supercluster 
and 13 in the Virgo cluster. The high sensitivity of our observations (rms noise $\sim 0.5$ mJy) resulted in the detection of
37 objects and significant upper limits were obtained for the remaining ones. \\
Including the present observations the \HI\ survey of the Coma Supercluster has reached virtually the completion
(94 \% among the late-type members).\\
Combining all data, we determine with high significance that the typical scale of \HI\
deficiency around the Coma cluster is 2-3 Mpc, i.e. one virial radius.\\
With the present data a meaningful determination of the HIMF can be obtained only for 
log$M_{HI}>9.0 M_\odot$, insufficient to compare with the deeper HIMF of Virgo (Paper I) and 
of isolated galaxies (Zwaan et al. 2003). Comparing cluster with non-cluster Supercluster
members we detect however a shortage of high \HI\ mass galaxies among cluster members that
can be ascribed to the pattern of \HI\ deficiency found in rich clusters.

\acknowledgements{ 
The Arecibo Observatory is part of the National Astronomy and Ionosphere Center, 
which is operated by Cornell University under a cooperative agreement with the 
National Science Foundation. This research also has made use of the Lyon-Meudon 
Extragalactic Database (LEDA), recently incorporated in HyperLeda and  the NASA/IPAC 
Extragalactic Database (NED) which is operated by the Jet Propulsion Laboratory, 
California Institute of Technology, under contract with the National Aeronautics and Space      
Administration, and the Goldmine database.  
}

\newpage

\begin{table*}
{\footnotesize
\begin{tabular}{llccccccccc}
\multicolumn{11}{l}{\footnotesize {\bf Table 3} Parameters of the newly observed galaxies.} \\
\smallskip \\
\hline
\vspace{-2 mm} \\
Obj. & RA & Dec & $V_{opt}$ & $\sigma$ & $S_p$ & $V_{HI}$ & $W_{50}$ & $W_{20}$ & $I_{HI}$ & Qual. \\ 
 & \multicolumn{2}{c}{J2000.0} & km/s & mJy/beam & mJy/beam & km/s & km/s & km/s & Jy km/s \\   
\hline
\vspace{-2 mm} \\
\multicolumn{10}{l}{Coma Supercluster} \\   
\vspace{-2 mm} \\
CGCG 97-078  & 114316.24 & 194455.6 & 7560$\pm$70  & 0.54 &   -  &   -          &  -  &  -  &   - &  - \\ 
CGCG 127-018 & 113944.62 & 224107.7 &         -    & 0.56 & 17.0 &  6922$\pm$2  & 147 & 168 & 2.09$\pm$0.06  & 1 \\ 
CGCG 127-039 & 114330.88 & 230043.3 & 6908$\pm$60  & 0.62 & 22.4 &  6911$\pm$1  &  37 &  58 & 0.82$\pm$0.04  & 2 \\ 
CGCG 97-124  & 114456.97 & 194353.9 & 7771$\pm$60  & 0.66 &   -  &   -   &   -   &  -  &  -   &  - \\ 
CGCG 127-055 & 114646.66 & 211616.9 & 6615$\pm$52  & 0.67 &  9.3 &  6626$\pm$3  & 215 & 242 & 1.86$\pm$0.08  & 1 \\ 
CGCG 127-067 & 115039.40 & 205426.1 & 6349$\pm$39  & 0.43 & 3.8  &  6400$\pm$16 & 283 & 304 & 0.73$\pm$0.10& 1  \\ 
CGCG 157-044 & 115123.19 & 264703.6 & 6624$\pm$44  & 0.64 &  5.1 &  6607$\pm$4  & 240 & 260 & 0.84$\pm$0.08  & 1 \\ 
CGCG 97-172  & 115214.30 & 183905.7 & 7650$\pm$56  & 0.36 & 1.6  &  7826$\pm$16 & 214 & 227 & 0.21$\pm$0.10& 3  \\ 
CGCG 127-137W & 120141.90 & 202417.3 & 6794         & 0.49 & 3.4  &  6871$\pm$16 & 357 & 398 & 1.07$\pm$0.10& 1  \\ 
CGCG 98-023  & 120144.40 & 175353.8 & 6905$\pm$48  & 0.38 & 5.8  &  6905$\pm$16 & 262 & 282 & 1.01$\pm$0.10& 1  \\ 
CGCG 127-136 & 120147.40 & 210506.7 & 6930         & 0.29 & 17.5 &  6675$\pm$16 & 173 & 228 & 2.36$\pm$0.10& 2  \\ 
CGCG 127-138 & 120155.47 & 204452.1 & 7192$\pm$46  & 0.60 &   -  &   -          &  -  &  -  &   - & -  \\ 
CGCG 128-015 & 120456.20 & 211427.5 & 6729$\pm$16  & 0.44 & 19.0 &  6741$\pm$16 & 102 & 124 & 1.64$\pm$0.10& 2  \\ 
CGCG 99-002  & 121755.86 & 182357.8 & 7640$\pm$59  & 0.90 &  2.7 &  (7605) &  - &  - &  -   & 4 \\ 
CGCG 128-072 & 121808.27 & 244118.5 & 6838$\pm$51  & 0.63 &  8.3 &  6795$\pm$6  & 119 & 174 & 0.82$\pm$0.07  & 1 \\ 
CGCG 99-013  & 121910.00 & 191626.9 & 7297$\pm$16  & 0.36 & 7.2  &  7336$\pm$16 & 235 & 251 & 1.41$\pm$0.10& 1  \\ 
CGCG 128-081 & 122052.50 & 252547.2 & 7112         & 0.42 & 5.0  &  7204$\pm$16 & 290 & 312 & 1.00$\pm$0.10& 1  \\ 
CGCG 129-004 & 122949.40 & 222219.4 & 6847$\pm$70  & 0.57 &  5.5 &  6736$\pm$7  & 275 & 317 & 1.09$\pm$0.08  & 1 \\ 
CGCG 159-048 & 124011.35 & 311038.1 & 7096$\pm$44  & 0.65 &  7.0 &  7064$\pm$8  & 309 & 363 & 1.60$\pm$0.10    & 1\\ 
CGCG 159-071 & 124543.41 & 292558.5 & 6936$\pm$44  & 1.06 & 17.1 &  6971$\pm$1  & 189 & 202 & 2.60$\pm$0.10    & 1\\ 
CGCG 159-097 & 125206.77 & 270134.3 & 6573$\pm$190 & 0.65 &  3.2 &  6424$\pm$30 & 260 & 357 & 0.60$\pm$0.10    & 3 \\ 
CGCG 160-009 & 125432.95 & 282234.7 & 7079$\pm$63  & 0.70 &   -  &   -          &  -  &  -  &   - & - \\ 
FOCA 636     & 125756.70 & 275930.0 & 4649         & 0.40 &  1.8 &  4605$\pm$30 & 130 & 219 & 0.18$\pm$0.05  & 3 \\ 
FOCA 610     & 125757.70 & 280342.0 & 8299         & 0.38 &  1.7 &  8125$\pm$8  & 211 & 236 & 0.19$\pm$0.05  & 3  \\ 
CGCG 130-003 & 125947.24 & 214845.7 & 7094$\pm$32  & 0.55 &  4.4 &  7140$\pm$4  & 335 & 357 & 0.51$\pm$0.08  & 3 \\ 
CGCG 160-261 & 130059.10 & 275358.6 & 6896$\pm$7   & 0.37 & -	 &	-	&   -  &  -   &	-	   & -	 \\ %
CGCG 160-128 & 130422.57 & 284838.5 & 8054$\pm$36  & 0.74 & 25.7 &  7920$\pm$1  & 115 & 137 & 2.46$\pm$0.07  & 1 \\ 
CGCG 160-138 & 130635.52 & 271006.2 & 7852$\pm$60  & 0.37 &   -  &   -          &  -  &  -  &   - &  - \\ 
CGCG 160-146 & 130814.00 & 273055.3 & 7337$\pm$23  & 0.25 & -	 &   -		&  -  &  -  &	- &  -	\\ %
CGCG 160-169 & 131440.60 & 295951.8 & 6960$\pm$60  & 0.69 &  6.2 &  6850$\pm$4  & 298 & 318 & 1.26$\pm$0.10   & 1 \\ 
CGCG 130-029 & 131730.80 & 203555.1 & 6671$\pm$71  & 0.36 & 4.6  &  6560$\pm$16 & 317 & 325 & 1.30$\pm$0.10& 1  \\ 
CGCG 160-195 & 131947.60 & 304931.9 & 7297$\pm$15  & 0.34 & 5.0  &  7247$\pm$16 & 158 & 223 & 0.63$\pm$0.10& 2  \\ 
CGCG 161-048 & 132557.14 & 313703.6 & 7268$\pm$44  & 0.66 &   -  &   -          &  -  &  -  &   - &  - \\ 
CGCG 161-051 & 132643.29 & 303026.7 & 7150$\pm$60  & 1.52 &   -  &   -          &  -  &  -  &   - &  - \\ 
CGCG 161-054 & 132703.07 & 305834.3 & 7668$\pm$44  & 0.62 &  7.0 &  6756$\pm$3  & 284 & 303 & 1.34$\pm$0.09  & 1 \\ 
\vspace{-2 mm} \\
\multicolumn{10}{l}{Virgo Cluster} \\   
\vspace{-2 mm} \\                          		        				      
VCC  30      & 121054.50 & 155654.6 &   -          & 0.92 &  9.5 & 2084 $\pm$16 & 116 & 148 & 0.83$\pm$0.10 & 2 \\ 
VCC  85      & 121336.50 & 130201.1 &   -          & 0.32 &  2.4 & 1932 $\pm$16 & 74  &  84 & 0.14$\pm$0.10 & 2 \\ 
VCC  113     & 121432.90 & 120612.0 & 2155         & 0.33 & 17.5 & 2091 $\pm$16 & 148 & 171 & 2.12$\pm$0.10 & 2 \\ 
VCC  137     & 121508.60 & 145819.5 &   -          & 0.56 &  -   &    (4000)	&  -  &  -  &	-    & 4 \\ 
VCC  429     & 122044.20 & 143803.3 &   -          & 0.39 &  4.0 & 600  $\pm$16 & 87  & 109 & 0.27$\pm$0.10 & 2 \\ 
VCC  578     & 122243.60 & 183252.1 &   -          & 0.39 &  4.0 & 6490 $\pm$16 & 161 & 178 & 0.48$\pm$0.10 & 1 \\ 
VCC  1574    & 123432.90 & 151052.3 &   -          & 0.17 &  2.3 & 639  $\pm$16 & 127 & 161 & 0.22$\pm$0.10 & 2 \\ 
VCC  1623    & 123531.90 & 163646.9 &   -          & 0.30 &  1.9 & 2108 $\pm$16 & 78  & 91  & 0.13$\pm$0.10 & 2 \\ 
VCC  1821    & 124008.90 & 065302.1 &   -          & 0.31 &  2.5 & 1007 $\pm$16 & 75  & 87  & 0.16$\pm$0.10 & 2 \\ 
VCC  1873    & 124118.60 & 063127.0 &   -          & 0.26 & 17.0 & 1692 $\pm$16 & 112 & 134 & 1.37$\pm$0.10 & 2 \\ 
VCC  1898    & 124157.50 & 034909.4 & 881 $\pm$23  & 0.50 &  -   &	-	&  -   &  -   &		-     & -\\ 
VCC  2071    & 124825.40 & 091856.8 &   -          & 0.32 &  2.7 & 6484 $\pm$16 & 185 & 196 & 0.40$\pm$0.10 & 2 \\ 
CGCG 43066   & 125515.40 & 025348.4 & 2798$\pm$23  & 0.47 & 21.0 & 2802 $\pm$16 & 382 & 403 & 6.09$\pm$0.10 & 1 \\ 
\vspace{-2 mm} \\
\hline
\label{obs_dat}
\end{tabular}
}
\end{table*}
   \onecolumn
\setcounter{table}{3}   
   \scriptsize{
   \begin{longtable}{lccccrcrc}
   \caption{Basic \HI\ properties of late-type members to the Coma Supercluster (including HL).}\\
   \hline
   \hline
   \noalign{\smallskip}
    CGCG & Type & $m_p$ & Cloud & V & log $M_{HI}$ & Ref. & $Def_{HI}$ & Qual \\  
             &      & mag &  &  \kms & $M\odot$  &      &            &  \\
   \noalign{\smallskip}
   \hline
   \noalign{\smallskip}
   \endfirsthead
   \caption{Continue}\\
   \hline
   \noalign{\smallskip}
   CGCG & Type & $m_p$ & Cloud & V & log $M_{HI}$ & Ref. & $Def_{HI}$ & Qual \\  
             &      & mag &  &  \kms & $M\odot$   &      &            &  \\
   \noalign{\smallskip}
   \hline
   \noalign{\smallskip}
   \endhead
   \hline
   \endfoot          
  097-005   &	  Sc &15.5& Isol. &   6129 &	     9.72  & 126&     -0.29 &	 1 \\
  097-026   &	 Pec &13.9&  Pair &   6202 &	     9.88  &  39&     -0.45 &	 1 \\
  097-027   &	  Sc &14.6&  Pair &   6630 &	     9.27  &  24&      0.28 &	 2 \\
  097-036   &	 S.. &15.7&  Pair &   6595 &	      -    &  - &	-   &	 - \\
  097-062   &	 Pec &15.5& A1367 &   7809 &	     9.33  &  40&      0.31 &	 1 \\
  097-063   &	 Pec &15.7& A1367 &   6102 &	     9.09  &  24&      0.22 &	 2 \\
  097-064   &	 S.. &15.6& A1367 &   5976 &	     9.17  & 126&      0.03 &	 3 \\
  097-068   &	 Sbc &14.7& A1367 &   5974 &	     9.99  & 126&     -0.14 &	 1 \\
  097-072   &	  Sa &15.0& A1367 &   6332 &	     9.14  & 126&      0.50 &	 1 \\
  097-073   &	 Pec &15.6& A1367 &   7290 &	     9.31  &  40&      0.16 &	 2 \\
  097-076   &	  Sb &15.5& A1367 &   7060 &	 $<$   8.39  &  40&  $>$   1.39 &	 - \\
  097-078   &	  Sa &15.2& A1367 &   7526 &	 $<$   8.68  & 193&  $>$   1.28 &	 - \\
  097-079   &	 Pec &15.7& A1367 &   6996 &	     9.21  &  40&      0.25 &	 1 \\
  097-082   &	  Sa &15.0& A1367 &   6100 &	 $<$   8.68  &   2&  $>$   1.01 &	 5 \\
  097-087   &	 Pec &14.3& A1367 &   6735 &	     9.83  &  40&      0.19 &	 1 \\
  097-091   &	  Sa &14.7& A1367 &   7368 &	     9.77  & 126&     -0.18 &	 1 \\
  097-092   &	 Sbc &15.5& A1367 &   6373 &	     9.18  &  24&      0.31 &	 4 \\
  097-093   &	 Pec &15.5& A1367 &   4857 &	     9.03  &   4&      0.57 &	 4 \\
  097-102N  &	  Sa &15.1& A1367 &   6361 &	     9.21  &  40&      0.36 &	 1 \\
  097-111S  &	 Pec &16.5& A1367 &   7239 &	      -    &  - &	-   &	 - \\
  097-114   &	 Pec &15.4& A1367 &   8257 &	     9.44  & 193&     -0.17 &	 1 \\
  097-119   &	  Sa &15.7& A1367 &   5256 &	     8.92  &   4&      0.22 &	 1 \\
  097-120   &	  Sa &14.5& A1367 &   5595 &	     8.80  &   4&      0.90 &	 4 \\
  097-121   &	 Sab &14.6& A1367 &   6571 &	     9.37  & 126&      0.28 &	 1 \\
  097-122   &	 Pec &14.9& A1367 &   5468 &	     9.35  & 126&      0.49 &	 1 \\
  097-129E  &	 Sbc &15.7& A1367 &   6009 &	      -    &  - &	-   &	 - \\
  097-129W  &	  Sb &14.0& A1367 &   5082 &	    10.09  &  24&      0.18 &	 1 \\
  097-130   &	  Sa &15.5& A1367 &   6697 &	 $<$   8.83  &   1&  $>$   0.40 &	 5 \\
  097-138   &	 Pec &15.5& A1367 &   5317 &	     9.69  &  40&     -0.22 &	 3 \\
  097-149   &	 S.. &15.6& A1367 &   6060 &	 $<$   9.25  & 105&  $>$   0.07 &	 - \\
  097-151   &	 Sab &15.6& Isol. &   5854 &	     9.05  &  88&      0.34 &	 1 \\
  097-152   &	  Sa &15.5& A1367 &   6166 &	     9.40  & 126&      0.23 &	 1 \\
  097-168   &	 S.. &15.7& Isol. &   5996 &	 $<$   8.91  &  88&  $>$   0.25 &	 - \\
  097-169   &	  Sc &15.7&  Pair &   5975 &	     9.45  &   6&      0.05 &	 2 \\
  097-172   &	 S.. &15.7&  Pair &   7826 &	     8.71  & 193&      0.38 &	 3 \\
  098-002   &	  Sb &15.6& Isol. &   6206 &	     9.41  &  88&      0.10 &	 1 \\
  098-007   &	 Sbc &15.5& Isol. &   6350 &	     9.86  &  88&     -0.30 &	 1 \\
  098-013   &	  Sc &15.1& Isol. &   6949 &	     9.65  &  40&     -0.12 &	 1 \\
  098-016   &	  Sc &15.3& Isol. &   6449 &	     9.72  &  15&      0.11 &	 1 \\
  098-017   &	 Sbc &15.7& Isol. &   7015 &	     9.66  &  40&      0.05 &	 2 \\
  098-023   &	  Sb &15.1& Tripl. &   6905 &	     9.30  & 193&     -0.01 &	 1 \\
  098-034   &	 S.. &14.8& N4065 G &	6432 &       9.07  &  40&      0.11 &	 3 \\
  098-041   &	  Sc &15.7& N4065 G &	7551 &       9.28  &  24&      0.65 &	 2 \\
  098-046   &	  Sa &14.3& N4065 G &	6220 &       9.58  &  24&      0.07 &	 1 \\
  098-050    &   Sc  &14.1 &IC202 HL  &  4372	  &  9.44  &  24   &   -0.04 &  1 \\  
  098-051    &   Sa  &14.6 &IC202 HL  &  4283	  &  9.6   &  88   &   -0.35 &  4 \\  
  098-058   &	 Sbc &14.7& Isol. &   7207 &	     9.95  &  24&      0.00 &	 1 \\
  098-067   &	 S.. &15.7& Isol. &   7610 &	 $<$   9.64  & 184&  $>$  -0.71 &	 - \\
  098-071   &	 Pec &15.5&  Pair &   6881 &	     9.62  & 184&     -0.05 &	 1 \\
  098-072   &	 S.. &15.7&  Pair &   7946 &	      -    &  - &	-   &	 - \\
  098-073   &	  Sb &15.7& Tripl. &   6440 &	      -    &  - &	-   &	 - \\
  098-074   &	  Sa &15.6& Isol. &   7471 &	 $<$   9.10  &  88&  $>$   0.36 &	 - \\
  098-081   &	  Sa &15.2&  Pair &   7177 &	      -    &  - &	-   &	 - \\
  098-085   &	  Sc &14.7& Tripl. &   7042 &	     9.69  &  24&     -0.13 &	 2 \\
  098-087   &	 S.. &15.3&  Pair &   7540 &	      -    &  - &	-   &	 - \\
  098-116   &	  Sc &14.9& Isol. &   6229 &	     9.91  &  24&     -0.34 &	 1 \\
  099-002   &	 S.. &15.5& Isol. &   7605 &	     9.02  & 193&     -0.08 &	 3 \\
  099-013   &	  Sc &15.7& Isol. &   7736 &	     9.50  & 193&     -0.40 &	 1 \\
  100-005   &	 Pec &14.4& Isol. &   6611 &	     9.37  &  24&      0.41 &	 1 \\
  100-012   &	 Pec &15.3& Isol. &   6481 &	     9.32  &  24&     -0.12 &	 2 \\
  101-033   &	  Sc &15.7& Isol. &   6729 &	     9.68  &  43&      0.07 &	 1 \\
  101-043   &	  Sa &15.0& Isol. &   6677 &	 $<$   8.74  &  88&  $>$   0.96 &	 - \\
  101-049   &	 Sbc &14.9& Isol. &   7148 &	    10.06  &   7&     -0.24 &	 2 \\
  101-054   &	 Sab &13.8& Isol. &   6606 &	     9.92  &  24&     -0.20 &	 3 \\
  127-005   &	 Sbc &15.4& Isol. &   6864 &	     9.64  &   6&      0.04 &	 2 \\
  127-018   &	  Sb &15.0& Isol. &   6922 &	     9.63  & 193&     -0.14 &	 1 \\
  127-025N  &	  Sc &15.3&  Pair &   7142 &	      -    &  - &	-   &	 - \\
  127-025S  &	 Sbc &14.5&  Pair &   7076 &	     9.87  & 126&      0.09 &	 1 \\
  127-026   &	 Sbc &14.8& Isol. &   6871 &	     9.95  &  24&     -0.17 &	 1 \\
  127-033   &	  Sc &15.2& Isol. &   6300 &	     9.77  &  24&      0.00 &	 1 \\
  127-035   &	  Sa &15.4& Isol. &   6817 &	     9.49  &  88&      0.15 &	 1 \\
  127-037   &	 Pec &15.4& Isol. &   6186 &	     9.54  &  24&     -0.10 &	 2 \\
  127-038   &	  Sc &14.0& Isol. &   6913 &	    10.36  &   2&     -0.20 &	 1 \\
  127-039   &	 Sbc &15.3& Isol. &   6911 &	     9.21  & 193&      0.25 &	 2 \\
  127-049   &	 Pec &15.5& A1367 &   7061 &	     9.34  &   6&      0.32 &	 1 \\
  127-050   &	 Sbc &14.8& N3937 G &	6752 &       9.95  &   2&      0.00 &	 1 \\
  127-051N  &	 Pec &15.8& A1367 &   7288 &	      -    &  - &	-   &	 - \\
  127-052   &	  Sa &14.0& A1367 &   6946 &	     9.73  & 126&      0.16 &	 1 \\
  127-053   &	 Sbc &15.4& Isol. &   6409 &	    1 -    &  24&     -0.11 &	 1 \\
  127-054   &	  Sb &14.2& N3937 G &	7026 &      10.38  &   2&      0.10 &	 1 \\
  127-055   &	 Pec &15.1& Isol. &   6626 &	     9.54  & 193&     -0.45 &	 1 \\
  127-056   &	  Sb &15.7& Isol. &   6814 &	     9.67  & 126&     -0.05 &	 1 \\
  127-061   &	  Sc &15.4& Isol. &   5954 &	     9.85  &  24&     -0.20 &	 1 \\
  127-067   &	 S.. &15.5& N3937 G &	6400 &       9.18  & 193&     -0.31 &	 1 \\
  127-071   &	 Pec &15.4& N3937 G &	6388 &       9.36  &  24&      0.13 &	 2 \\
  127-072   &	  Sc &14.6& N3937 G &	6438 &       9.89  &   3&     -0.05 &	 1 \\
  127-073   &	  Sb &15.1& N3937 G &	6439 &       9.01  &   1&      0.81 &	 4 \\
  127-082   &	  Sc &14.7& N3937 G &	6654 &       9.47  &   2&      0.02 &	 2 \\
  127-083   &	 Sbc &15.1& N3937 G &	6743 &       9.39  &   6&      0.06 &	 3 \\
  127-085   &	  Sa &15.5& N3937 G &	6595 &   $<$   8.99  &  88&  $>$   0.46 &	 - \\
  127-087   &	Sbc  &15.4 &N4005 HL  &  4941	  &  9.66  &  126  &   0.05  &  1 \\  
  127-095   &	  Sc &14.2& N3937 G &	6199 &      10.04  &   2&      0.03 &	 1 \\
  127-099   &	  Sc &14.5&  Pair &   6458 &	     9.71  & 126&      0.10 &	 1 \\
  127-100   &	  Sb &14.9& N3937 G &	6854 &       9.58  &   2&      0.06 &	 3 \\
  127-104   &	 Sbc &15.5& Isol. &   6814 &	     9.63  &   6&      0.19 &	 1 \\
  127-106   &	Sb   &14.5 &N4005 HL  &  5027	  &  9.88  &  24   &   -0.32 &  1 \\  
  127-107   &	 Sbc &15.7& Isol. &   6355 &	     9.36  & 126&      0.02 &	 3 \\
  127-109   &	Sbc  &15.7 &N4005 HL  &  4731	  &  9.4   &  32   &   -0.04 &  2 \\  
  127-110   &	Sbc  &14.4 &N4005 HL  &  4495	  &  9.81  &  24   &   0.22  &  1 \\  
  127-111   &	Sbc  &15.7 &N4005 HL  &  4617	  &  9.47  &  111  &   -0.41 &  1 \\  
  127-112   &	Sbc  &14.8 &N4005 HL  &  4828	  &  9.63  &  32   &   0.25  &  1 \\  
  127-114E  &	 Pec &15.0 &N4005 HL  &  4771	  &  9.84  &  24   &   -0.02 &  3 \\  
  127-114W  &	 Pec &15.0 &N4005 HL  &  4771	  &  9.84  &  24   &   -0.13 &  3 \\  
  127-118   &	 Sc  &15.2 &N4005 HL  &  4551	  &  9.38  &  24   &   0.13  &  1 \\  
  127-120   &	Sb   &14.1 &N4005 HL  &  4470	  &  9.32  &  24   &   0.22  &  3 \\  
  127-121   &	Sb   &15.7 &N4005 HL  &  4131	  &   -    &   -   &	 -   &  - \\  
  127-123   &	 Sc  &14.7 &N4005 HL  &  4479	  &  9.84  &  24   &   -0.03 &  1 \\  
  127-127   &	Sb   &14.6 &N4005 HL  &  4048	  &  9.34  &  111  &   0.14  &  1 \\  
  127-133   &	 Sc  &15.3 &N4005 HL  &  4667	  &  9.12  &  32   &   -0.08 &  1 \\  
  127-136   &	 S.. &15.7&  Pair &   6675 &	     9.68  & 193&     -1.02 &	 2 \\
  127-137W  &	 S.. &16.0& N4065 G &	6871 &       9.38  & 193&     -0.32 &	 1 \\
  127-138   &	 S.. &15.5& Isol. &   7210 &	 $<$   8.77  & 193&  $>$  -0.09 &	 - \\
  127-139   &	  Sa &15.5& Isol. &   6713 &	     9.29  &  24&      0.34 &	 3 \\
  128-002   &	 S.. &15.7& N4065 G &	6780 &        -    &  - &	-   &	 - \\
  128-003   &	 Pec &14.6& Isol. &   6435 &	     9.69  &   2&     -0.11 &	 2 \\
  128-015   &	  Sb &15.3&  Pair &   6741 &	     9.51  & 193&     -0.11 &	 2 \\
  128-016   &	 S.. &15.2& Isol. &   6619 &	     9.30  &  24&     -0.31 &	 1 \\
  128-021   &	 Sbc &15.4& Isol. &   7064 &	     9.83  &  24&      0.02 &	 1 \\
  128-023   &	  Sa &14.4& N4065 G &	6719 &       9.92  &   2&     -0.30 &	 1 \\
  128-029E  &	 S.. &16.7& IC762 G &	6634 &        -    &  - &	-   &	 - \\
  128-029W  &	 S.. &16.2& IC762 G &	7158 &       9.39  &  40&     -0.10 &	 3 \\
  128-031N  &	 S.. &16.5& Isol. &   6936 &	      -    &  - &	-   &	 - \\
  128-037   &	 Sbc &14.8& IC762 G &	7194 &       9.53  &   2&      0.11 &	 2 \\
  128-042N  &	 S.. &16.3& Isol. &   7319 &	     9.78  &  43&     -0.30 &	 3 \\
  128-044   &	 S.. &15.7& Isol. &   6922 &	    10.25  &   7&     -0.36 &	 1 \\
  128-049   &	  Sc &15.0& Isol. &   6445 &	     9.21  &  24&      0.49 &	 4 \\
  128-052   &	  Sb &15.6& Isol. &   6692 &	     9.48  &  22&      0.07 &	 2 \\
  128-053   &	 Sbc &15.6& Isol. &   7307 &	    10.06  &  40&      0.01 &	 1 \\
  128-057   &	 S.. &15.6& Isol. &   6998 &	 $<$   9.51  & 184&  $>$  -0.35 &	 - \\
  128-058   &	 S.. &15.7& N4213 G &	6778 &       9.52  &   7&     -0.25 &	 2 \\
  128-059   &	  Sb &15.6& N4213 G &	6228 &       9.95  &   7&     -0.28 &	 1 \\
  128-063   &	  Sa &15.3&  Pair &   6747 &	     9.97  &  24&      0.06 &	 1 \\
  128-066   &	 S.. &15.1& N4213 G &	6526 &   $<$   9.11  &  88&  $>$   0.34 &	 - \\
  128-069   &	 Sbc &15.6& N4213 G &	7188 &       9.16  & 111&      0.28 &	 2 \\
  128-072   &	 Pec &15.4& Isol. &   6795 &	     9.21  & 193&      0.02 &	 1 \\
  128-073   &	  Sb &14.7& Isol. &   6948 &	     9.96  & 126&      0.01 &	 1 \\
  128-075   &	  Sc &15.5& Isol. &   6682 &	     9.93  & 126&      0.10 &	 3 \\
  128-079   &	  Sc &15.6& Isol. &   6630 &	     9.71  &  40&     -0.25 &	 2 \\
  128-080   &	  Sb &15.0& Isol. &   7349 &	     9.38  &  24&      0.05 &	 3 \\
  128-081W  &	 S.. &16.5& Isol. &   7204 &	     9.33  & 193&     -0.29 &	 1 \\
  128-082   &	  Sb &15.7& Isol. &   6910 &	    1 -    & 111&     -0.05 &	 1 \\
  128-087   &	  Sc &15.3& Isol. &   6671 &	     9.55  &   5&      0.13 &	 1 \\
  128-089   &	  Sa &14.2& Isol. &   6841 &	     9.23  &  40&      0.35 &	 3 \\
  128-090   &	  Sc &15.5& Isol. &   6776 &	     9.87  &   7&     -0.35 &	 1 \\
  129-004   &	 S.. &15.2& Isol. &   6736 &	     9.33  & 193&     -0.08 &	 1 \\
  129-009   &	  Sa &15.3&  Pair &   6415 &	     9.27  &  24&      0.11 &	 1 \\
  129-013   &	 S.. &15.7& Isol. &   6962 &	     9.53  & 184&     -0.03 &	 1 \\
  129-016   &	 S.. &15.5& N4615 HL &	4979 &        -    &  - &	-   &	 - \\
  129-018   &	  Sc &13.8& N4615 HL &	4716 &      10.05  &  24&     -0.23 &	 1 \\
  129-020   &	  Sb &14.8& Isol. &   6579 &	     9.50  &  15&      0.14 &	 3 \\
  129-021   &	 S.. &15.3& Isol. &   6697 &	     9.85  &  24&     -1.15 &	 2 \\
  129-022   &	 Sab &14.4& Isol. &   6972 &	     9.98  &   2&     -0.25 &	 3 \\
  129-023   &	 S.. &15.7& Isol. &   6746 &	     9.41  & 184&     -0.42 &	 4 \\
  129-025   &	  Sc &13.5& N4615 HL &	4380 &      10.02  &   6&      0.12 &	 1 \\
  130-002   &	 S.. &15.6& Isol. &   6663 &	     9.61  & 184&     -0.64 &	 1 \\
  130-003   &	  Sb &15.4& Isol. &   7140 &	     9.04  & 193&      0.68 &	 3 \\
  130-005   &	 Sbc &15.5& Isol. &   7058 &	     9.30  & 127&      0.21 &	 1 \\
  130-006   &	 Sbc &15.0& Isol. &   6521 &	     9.33  &  24&      0.04 &	 2 \\
  130-008   &	  Sc &14.9& Isol. &   7266 &	     9.73  &   2&     -0.53 &	 2 \\
  130-009   &	 Sbc &15.3&  Pair &   6335 &	     9.93  &  24&     -0.03 &	 1 \\
  130-012   &	 Sbc &15.2&  Pair &   7131 &	     9.97  &  24&      0.19 &	 1 \\
  130-014   &	 Sbc &15.1& Isol. &   7096 &	     9.57  & 126&      0.09 &	 1 \\
  130-021   &	  Sa &15.4& Isol. &   7163 &	     9.33  & 127&      0.18 &	 1 \\
  130-025   &	  Sa &15.5& Isol. &   7001 &	     9.68  &  15&      0.02 &	 1 \\
  130-026   &	  Sc &15.5& Quadr. &   6870 &	     9.85  &  24&     -0.03 &	 1 \\
  130-027   &	 Sbc &15.6& Quadr. &   6834 &	     9.94  &  40&     -0.13 &	 1 \\
  130-029   &	  Sc &15.4&  Pair &   6560 &	     9.39  & 193&      0.04 &	 1 \\
  131-008   &	 Sbc &15.6& Isol. &   5972 &	     9.76  &  88&     -0.04 &	 1 \\
  131-009   &	  Sc &15.3& Isol. &   7522 &	     9.64  &  24&     -0.04 &	 1 \\
  157-012   &	 Sbc &15.1& Isol. &   6814 &	     9.73  &  40&     -0.19 &	 1 \\
  157-032   &	  Sa &15.2& Isol. &   6811 &	     9.16  &  24&      0.64 &	 4 \\
  157-035   &	  Sb &13.7& Isol. &   6281 &	    10.18  &  24&     -0.05 &	 1 \\
  157-044   &	 Pec &15.4& Isol. &   6607 &	     9.19  & 193&      0.22 &	 1 \\
  157-051   &	 Sc  &15.3 &N4005 HL  &  5151	  &  9.65  &  24   &   -0.04 &  2 \\  
  157-062   &	 Pec &15.5& Isol. &   6882 &	     9.68  &   5&      0.01 &	 2 \\
  157-064   &	  Sb &14.8& Isol. &   6407 &	     9.84  &  40&      0.02 &	 2 \\
  157-075   &	  Sc &15.7& Isol. &   6694 &	     9.54  &   7&      0.06 &	 2 \\
  157-077   &	 S.. &15.4 &N4005 HL  &  4100	  &   -    &   -   &	 -   &  - \\  
  158-009   &	  Sb &14.0&  Pair &   7494 &	     9.07  &  24&      0.65 &	 4 \\
  158-010   &	 Sbc &15.2&  Pair &   7930 &	     9.46  &  24&      0.17 &	 2 \\
  158-029   &	 S.. &14.1 &N4169 HL  &  3836	  &  8.98  &  22   &   0.46  &  2 \\  
  158-030   &	Sab  &14.6 &N4169 HL  &  3970	  &  9.51  &  22   &   -0.35 &  2 \\  
  158-031   &	Sb   &13.8 &N4169 HL  &  3825	  &  10.03 &  39   &   -0.13 &  1 \\  
  158-036   &	  Sb &13.8& Isol. &   6532 &	     9.96  &   2&     -0.20 &	 1 \\
  158-038   &	 Sab &15.3& Isol. &   6725 &	     9.36  &  40&      0.07 &	 1 \\
  158-042   &	 S.. &14.8 &N4169 HL  &  3868	  &  8.97  &  88   &   0.11  &  2 \\  
  158-046   &	 S.. &15.0 &N4169 HL  &  3848	  &   -    &   -   &	 -   &  - \\  
  158-047   &	Sb   &13.5 &N4169 HL  &  3903	  &  9.79  &  39   &   0.07  &  1 \\  
  158-053N  &	  Sa &14.7&  Pair &   6599 &	     9.79  &   2&     -0.02 &	 1 \\
  158-054   &	 Pec &14.6& Isol. &   7685 &	     9.76  &   2&     -0.13 &	 3 \\
  158-055   &	  Sb &15.3& Isol. &   7650 &	     9.53  &  24&      0.46 &	 4 \\
  158-056   &	  Sa &15.5& Tripl. &   8102 &	     9.53  &  24&      0.16 &	 1 \\
  158-061   &	 Sa  &13.7 &N4169 HL  &  3876	  & $<$8.24  &  168  &   0.84  &  - \\  
  158-070   &	 Sbc &15.3& Quadr. &   7634 &	     9.58  &   5&      0.23 &	 1 \\
  158-081   &	 Pec &14.5& Isol. &   6734 &	     9.26  &   2&     -0.03 &	 2 \\
  158-091   &	 Sab &15.7& IC3165 G&	7607 &       9.54  &   2&     -0.17 &	 2 \\
  158-102   &	 S.. &15.7& N4615 HL &	4515 &       9.49  &   7&      0.03 &	 1 \\
  158-105   &	 Sbc &15.1& Isol. &   6824 &	     9.98  &  24&     -0.16 &	 1 \\
  158-112   &	 Sbc &14.4&  Pair &   7165 &	     9.73  &   2&      0.41 &	 4 \\
  159-004   &	 S.. &15.7&  Pair &   7004 &	     9.11  &  88&      0.15 &	 2 \\
  159-005   &	 Sbc &14.7&  Pair &   6996 &	     9.61  &   2&      0.27 &	 2 \\
  159-008   &	  Sb &14.6& Isol. &   7393 &	    10.03  & 126&      0.06 &	 1 \\
  159-009   &	 Sab &14.1& N4615 HL &	4551 &      10.14  &  24&     -0.52 &	 1 \\
  159-010   &	  Sb &15.7& Isol. &   7009 &	     9.62  &   5&      0.24 &	 1 \\
  159-019   &	 Sbc &14.9& N4615 HL &	4573 &       9.62  &   6&     -0.09 &	 1 \\
  159-031   &	  Sa &15.3&  Pair &   7511 &	     9.91  & 127&     -0.16 &	 1 \\
  159-033   &	  Sa &15.0& Isol. &   7674 &	     9.34  & 126&      0.54 &	 1 \\
  159-037   &	 Sab &14.6& Isol. &   7291 &	     9.73  & 126&     -0.28 &	 1 \\
  159-040   &	  Sa &15.2& Isol. &   7019 &	     9.93  &  24&     -0.34 &	 1 \\
  159-048   &	 S.. &15.5& Isol. &   7064 &	     9.53  & 193&     -0.25 &	 1 \\
  159-049S  &	 S.. &15.7&  Pair &   6330 &	      -    &  - &	-   &	 - \\
  159-054   &	  Sc &15.5& N4615 HL &	4759 &       9.41  &  88&     -0.01 &	 1 \\
  159-055   &	 Sbc &15.6&  Pair &   7737 &	    10.01  &  24&      0.01 &	 1 \\
  159-058   &	  Sa &15.5& Isol. &   6797 &	 $<$   9.50  & 184&  $>$  -0.07 &	 - \\
  159-059   &	 Sab &14.5& Isol. &   7528 &	     9.76  &   2&     -0.31 &	 2 \\
  159-060   &	 Pec &15.5& Isol. &   7182 &	     9.67  &   5&      0.04 &	 3 \\
  159-061   &	 Sbc &14.8& Isol. &   6966 &	     9.49  &   2&      0.28 &	 3 \\
  159-064   &	 S.. &15.6& Isol. &   7265 &	 $<$   9.64  & 184&  $>$  -0.12 &	 - \\
  159-068   &	  Sa &15.7& Isol. &   6313 &	     9.25  & 184&      0.03 &	 4 \\
  159-071   &	  Sc &15.5& Isol. &   6971 &	     9.72  & 193&     -0.03 &	 1 \\
  159-072N  &	 Pec &14.8&  Pair &   6631 &	    10.09  &   2&     -0.03 &	 4 \\
  159-072S  &	 Pec &14.8&  Pair &   6590 &	    10.03  &   2&     -0.12 &	 4 \\
  159-076   &	 Sbc &14.5& Isol. &   6743 &	     9.51  &   2&      0.39 &	 1 \\
  159-080   &	  Sb &15.7& Isol. &   6859 &	     9.54  & 126&      0.14 &	 1 \\
  159-081   &	 Sbc &15.5&  Pair &   8116 &	     9.85  &  40&     -0.20 &	 1 \\
  159-082   &	  Sc &14.8&  Pair &   8078 &	    10.01  & 126&     -0.18 &	 1 \\
  159-090   &	  Sc &15.5& Tripl. &   8315 &	    10.16  & 126&     -0.56 &	 1 \\
  159-091   &	 S.. &15.1& Isol. &   6443 &	     9.16  & 126&      0.00 &	 1 \\
  159-092   &	  Sc &14.9& N4615 HL &	4754 &       9.94  & 126&      0.02 &	 1 \\
  159-093   &	  Sc &15.3& N4615 HL &	5446 &   $<$   8.58  &  40&  $>$   0.63 &	 5 \\
  159-095   &	 Sbc &14.9& Isol. &   6837 &	     9.64  &   2&     -0.24 &	 1 \\
  159-096   &	  Sc &15.1& Isol. &   6186 &	     9.96  & 126&     -0.01 &	 1 \\
  159-097   &	 Pec &15.4& Isol. &   6424 &	     9.02  & 193&      0.21 &	 3 \\
  159-101   &	 Pec &15.3&  Coma &   7745 &	     9.12  &   6&      0.07 &	 1 \\
  159-102   &	 Sab &14.5&  Coma &   7061 &	     9.97  & 126&     -0.25 &	 3 \\
  160-001   &	  Sb &15.6&  Coma &   7945 &	     9.44  &  88&      0.09 &	 1 \\
  160-005   &	  Sb &14.8& Isol. &   6319 &	    10.06  & 126&     -0.01 &	 1 \\
  160-007   &	 S.. &15.4&  Coma &   6462 &	 $<$   8.82  &  88&  $>$   0.74 &	 - \\
  160-009   &	 S.. &15.5&  Coma &   7132 &	 $<$   8.84  & 193&  $>$   0.55 &	 - \\
  160-012   &	 S.. &15.7& Isol. &   6348 &	 $<$   9.36  & 184&  $>$  -0.16 &	 - \\
  160-015   &	 S.. &15.5&  Coma &   7443 &	 $<$   8.90  &  40&  $>$   0.63 &	 5 \\
  160-018   &	 S.. &15.3&  Coma &   7092 &	 $<$   8.82  &  88&  $>$   0.63 &	 - \\
  160-020   &	 Pec &15.5&  Coma &   4968 &	     8.93  &  24&      0.27 &	 4 \\
  160-025   &	  Sa &14.0&  Coma &   6702 &	 $<$   8.75  &  27&  $>$   0.92 &	 - \\
  160-026   &	  Sc &15.5&  Coma &   7545 &	     9.29  &  40&      0.23 &	 2 \\
  160-031   &	 S.. &15.7&  Coma &   6852 &	 $<$   8.84  &   6&  $>$   0.62 &	 5 \\
  160-032   &	  Sb &14.9&  Coma &   7747 &	 $<$   8.83  &   2&  $>$   0.76 &	 5 \\
  160-055   &	 Sab &14.2&  Coma &   7164 &	     9.34  &  40&      0.49 &	 2 \\
  160-058   &	 Sbc &15.5&  Coma &   7616 &	     9.49  & 126&      0.40 &	 1 \\
  160-064   &	 Pec &15.4&  Coma &   7368 &	 $<$   8.32  & 132&  $>$   0.93 &	 - \\
  160-067   &	 Pec &15.4&  Coma &   7655 &	     9.34  & 181&     -0.01 &	 2 \\
  160-073   &	 Pec &15.1&  Coma &   5425 &	     8.57  & 132&      0.96 &	 5 \\
  160-076   &	  Sc &15.6&  Coma &   5390 &	     9.64  & 132&     -0.35 &	 5 \\
  160-081   &	  Sb &14.7&  Coma &   5898 &	 $<$   8.86  &   1&  $>$   1.29 &	 5 \\
  160-086   &	 Pec &15.4&  Coma &   7499 &	     8.74  & 126&      0.76 &	 1 \\
  160-088   &	  Sb &14.6&  Coma &   7287 &	     9.35  &  40&      0.42 &	 1 \\
  160-095   &	  Sb &13.7&  Coma &   5482 &	     9.33  & 126&      0.96 &	 1 \\
  160-096N  &	 Pec &15.2&  Coma &   6892 &	     8.42  & 132&      0.95 &	 5 \\
  160-098   &	 Pec &15.3&  Coma &   8762 &	     9.05  &   1&      0.41 &	 2 \\
  160-102   &	 Sab &14.8&  Coma &   7095 &	     9.97  & 126&     -0.01 &	 1 \\
  160-106   &	 Pec &15.1&  Coma &   6876 &	     9.05  & 132&      0.54 &	 5 \\
  160-108   &	 Pec &15.5&  Coma &   8323 &	 $<$   8.29  &   6&  $>$   0.92 &	 5 \\
  160-114   &	 S.. &15.6&  Coma &   7454 &	 $<$   8.74  &  88&  $>$   0.81 &	 - \\
  160-121   &	  Sb &15.5&  Coma &   6676 &	     9.92  & 126&      0.02 &	 1 \\
  160-127   &	  Sc &15.5&  Coma &   5500 &	     9.71  & 126&     -0.10 &	 1 \\
  160-128   &	 Pec &15.3&  Coma &   7920 &	     9.73  & 193&     -0.33 &	 1 \\
  160-137   &	  Sa &13.9&  Coma &   7050 &	     9.79  &  40&     -0.05 &	 1 \\
  160-138   &	 S.. &15.7&  Coma &   7861 &	 $<$   8.56  & 193&  $>$   0.71 &	 - \\
  160-139   &	 Pec &15.0&  Coma &   4749 &	     9.96  & 126&     -0.19 &	 2 \\
  160-141   &	 Pec &15.5&  Coma &   7292 &	     8.95  &   1&      0.32 &	 1 \\
  160-146   &	  Sa &15.4&  Pair &   7385 &	 $<$   8.41  & 193&  $>$   1.12 &	 - \\
  160-148   &	  Sa &14.3& N5056 HL &	5988 &       9.75  & 126&     -0.07 &	 1 \\
  160-151   &	 Pec &15.1& N5056 HL &	6258 &       8.97  &  24&      0.31 &	 2 \\
  160-152W  &	  Sb &14.0& N5056 HL &	5610 &       9.95  &   2&      0.05 &	 1 \\
  160-155   &	  Sb &15.3& N5056 HL &	6366 &       9.60  & 126&      0.04 &	 1 \\
  160-156   &	  Sa &15.3&  Pair &   7262 &	     9.59  &  24&      0.28 &	 1 \\
  160-163N  &	 S.. &15.7& Isol. &   6877 &	 $<$   8.92  &  24&  $>$   0.28 &	 - \\
  160-164   &	  Sc &15.2& N5056 HL &	6074 &       9.69  & 126&     -0.19 &	 1 \\
  160-166   &	  Sb &13.6& N5056 HL &	6408 &       9.80  &   2&      0.24 &	 1 \\
  160-167   &	  Sb &15.0& N5056 HL &	6039 &       9.75  & 126&      0.02 &	 1 \\
  160-168   &	  Sc &14.2& Isol. &   7476 &	    10.36  &   2&     -0.40 &	 1 \\
  160-169   &	 S.. &15.6& Isol. &   6850 &	     9.41  & 193&      0.08 &	 1 \\
  160-173   &	  Sc &13.6& N5056 HL &	5592 &      10.12  & 126&     -0.12 &	 1 \\
  160-175   &	 S.. &15.1& N5056 HL &	5661 &       9.25  &  24&      0.18 &	 1 \\
  160-180   &	 S.. &15.4& N5056 HL &	5581 &        -    &  - &	-   &	 - \\
  160-181   &	  Sc &14.3& N5056 HL &	5550 &       9.87  & 126&     -0.11 &	 1 \\
  160-182   &	 Sab &15.0& Isol. &   6994 &	     9.62  &   5&      0.07 &	 1 \\
  160-183   &	 Pec &14.7& N5056 HL &	5605 &       9.57  &  40&     -0.14 &	 3 \\
  160-192   &	  Sb &14.3& Isol. &   6649 &	    10.25  & 126&     -0.09 &	 1 \\
  160-195   &	 S.. &15.7& Tripl. &   7247 &	     9.14  & 193&     -0.04 &	 2 \\
  160-206   &	 S.. &15.6& N5056 HL &	5053 &       8.99  &  40&      0.25 &	 1 \\
  160-207   &	  Sc &15.3& N5056 HL &	5081 &       9.71  &   5&      0.05 &	 1 \\
  160-209   &	 Pec &15.4&  Pair &   7168 &	     9.28  &  24&      0.28 &	 2 \\
  160-212   &	  Sa &14.9&  Coma &   7549 &	     8.78  & 132&      0.90 &	 5 \\
  160-213   &	 Pec &15.5&  Coma &   9386 &	 $<$   8.17  &   1&  $>$   1.11 &	 5 \\
  160-243   &	 S.. &15.6&  Coma &   5121 &	      -    &  - &	-   &	 - \\
  160-252   &	 Pec &15.1&  Coma &   7718 &	     9.01  &  40&      0.56 &	 1 \\
  160-257   &	  Sa &14.6&  Coma &   5821 &	     8.66  & 132&      0.97 &	 5 \\
  160-260   &	  Sa &13.7&  Coma &   7985 &	     9.18  & 126&      0.81 &	 1 \\
  160-261   &	 S.. &15.6&  Coma &   6917 &	 $<$   8.56  & 193&  $>$   0.81 &	 - \\
  161-029   &	  Sb &15.7& N5056 HL &	4930 &        -    &  - &	-   &	 - \\
  161-031   &	 Sbc &14.9& Tripl. &   7270 &	     9.75  &  40&      0.08 &	 2 \\
  161-040   &	  Sc &15.6& Isol. &   7260 &	     9.43  &  40&     -0.02 &	 1 \\
  161-041   &	 S.. &15.5& N5056 HL &	4979 &       8.81  &  88&      0.32 &	 1 \\
  161-043   &	  Sa &14.4& Isol. &   6638 &	     9.62  &  24&      0.33 &	 2 \\
  161-044   &	  Sc &15.3& N5056 HL &	4983 &       9.21  &  40&      0.48 &	 1 \\
  161-048   &	  Sa &15.1& Isol. &   7280 &	 $<$   8.82  & 193&  $>$   0.72 &	 - \\
  161-051   &	 S.. &15.6& Isol. &   7150 &	 $<$   9.17  & 193&  $>$  -0.12 &	 - \\
  161-052   &	 Pec &15.1& Isol. &   7072 &	     9.46  &  88&     -0.50 &	 2 \\
  161-054   &	  Sa &15.5& Isol. &   6756 &	     9.41  & 193&     -0.07 &	 1 \\
  161-063   &	 Sbc &15.5& Isol. &   7300 &	     9.97  &  88&     -0.06 &	 1 \\
  161-066   &	 S.. &15.7& Isol. &   7380 &	    10.12  &  40&     -0.56 &	 3 \\
  161-069   &	  Sb &14.6& Isol. &   7172 &	     9.88  &  24&     -0.27 &	 1 \\
  161-071   &	 Pec &14.9& N5056 L &	4827 &       9.84  &   5&     -0.26 &	 1 \\
  161-073   &	  Sb &14.2& Isol. &   7320 &	    10.38  &  24&     -0.21 &	 1 \\
   \noalign{\smallskip}
   \hline
   \label{sample_dat}
   \end{longtable}
}
1  : Giovanelli \& Haynes (1985); 
2  : Chincarini et al. (1983a); 
3  : Sullivan et al. (1981);
4  : Chincarini et al. (1983b); 
5  : Fontanelli (1984);
6  : Bothun et al. (1985);
7  : Williams \& Kerr (1981); 
15 : Haynes \& Giovanelli (1984); 
22 : Sulentic \&  Arp H (1982);
24 : Gavazzi (1987)
27 : Eder et al. (1991);
39 : Lewis et al. (1985);
40 : Gavazzi (1989);
43 : Salzer et al. (1990);
88 : Scodeggio \&  Gavazzi (1993);
105: Lu et al. (1993);
111: Mould et al. (1995);
126: Haynes et al. (1997);
127: Dell'Antonio et al. (1996);
132: Bravo Alfaro (2001);
168: Magri et al. (1994);
181: Beijsbergen (2003);
184: van Driel et al. (2000);
185: Vogt et al. (2004);
193: This work

\section{Appendix A: Notes on individual galaxies}

{\sf CGCG 127-018:} Our \HI\ detection (\VHI=6922$\pm$2 \kms, \mbox{$W_{50}$}=147 \kms\ and \IHI=2.1$\pm$0.06 \Jykms)
is consistent with our previous (van Driel et al. 2000) \nan\ detection 
(\VHI=6936$\pm$16 \kms, \mbox{$W_{50}$}=131 \kms\ and \IHI=2.7$\pm$0.3 \Jykms) and with the Arecibo detection
of Gavazzi (1987) at \VHI=6935 \kms. 

{\sf CGCG 127-039:} The integrated line intensity of our \HI\ detection 
(\VHI=6911$\pm$1 \kms, \mbox{$W_{50}$}=37 \kms\ and \IHI=0.82$\pm$0.04 \Jykms) is 3.4 times lower than that 
of our previous \nan\ detection (van Driel et al. 2000), with 
\VHI=6919$\pm$9 \kms, \mbox{$W_{50}$}=40 \kms\ and \IHI=2.8$\pm$0.2 \Jykms. 
The latter was made with an elongated \am{3}{6}$\times$21$'$ ($\alpha$$\times$$\delta$)HPBW, 
which is considerably larger than the \am{3}{6} round Arecibo beam. 
This difference can be due to the 13.6 $B$ mag SBbc spiral
NGC 3832,  \am{17}{2} due south of the target galaxy, whose mean \HI\ line parameters
(\VHI=6909$\pm$6 \kms, \mbox{$W_{50}$}=171 \kms\ and \IHI=10.4 \Jykms) are based on 5 spectra, all obtained at 
Arecibo (Chincarini et al. 1983a; Giovanardi \& Salpeter 1985; Lewis 1985; Lewis et al. 1985; 
Sullivan et al. 1981).

{\sf CGCG 127-055:} Our \HI\ detection (\VHI=6626$\pm$3 \kms, \mbox{$W_{50}$}=215 \kms\ and \IHI=1.9$\pm$0.08 \Jykms) 
is consistent with our previous measurement (van Driel et al. 2000) made at \nan, with
\VHI=6656$\pm$21 \kms, \mbox{$W_{50}$}=183 \kms\ and \IHI=2.1$\pm$0.3 \Jykms. 
Our new data have a 3.6 times better rms noise level.

{\sf CGCG 128-072:} Given the uncertainties, the global parameters of our \HI\ detection 
(\VHI=6795$\pm$6 \kms, \mbox{$W_{50}$}=119 \kms\ and \IHI=0.82$\pm$0.07 \Jykms)
are consistent with those of our previous \nan\ detection (van Driel et al. 2000), with 
\VHI=6848$\pm$81 \kms, \mbox{$W_{50}$}=213 \kms\ and \IHI=1.4$\pm$0.6 \Jykms.
Our new data have a 5.4 times better rms noise level.

{\sf CGCG 129-004:} discrepant values for the optical redshift have been published, 
4847 and 6729 \kms\ (Gavazzi et al. 1999a, and the compilation by Falco et al. 1999); 
we find an \HI\ value of 6736$\pm$7 \kms, consistent with the latter value.

{\sf CGCG 130-003:} there are two discrepant literature values for its optical redshift, 
7094 and 22,425 \kms\ (Straus et al. 1992; Gregory et al. 1988);
we find an \HI\ value of 7140$\pm$4 \kms, consistent with the former value.

{\sf CGCG 157-044:} our \HI\ detection (\VHI=6607$\pm$4 \kms, \mbox{$W_{50}$}=240 \kms\ and \IHI=0.84$\pm$0.08 \Jykms)
has a 1.7 times lower line intensity than our previous \nan\ detection (van Driel et al. 2000), 
with \VHI=6628$\pm$36 \kms, \mbox{$W_{50}$}=309 \kms\ and \IHI=1.5$\pm$0.4 \Jykms. The difference is less than
two times the uncertainty in the latter value, however, and therefore not significant.

{\sf CGCG 159-071:} the global parameters of our \HI\ detection (\VHI=6971$\pm$1 \kms\ and \mbox{$W_{50}$}=189 \kms\ 
and  \IHI=2.6$\pm$0.1 \Jykms) are consistent with those of our previous \nan\ detection
(van Driel et al. 2000), with \VHI=6985$\pm$6 \kms, \mbox{$W_{50}$}=164 \kms\ and \IHI=2.4$\pm$0.4 \Jykms.  

{\sf CGCG 159-097:} Its optical redshift, 6573$\pm$190 \kms, is not well determined.
We measured an \HI\ value of 6424$\pm$30 \kms, consistent with 3 of the published optical values -- 
only the optical velocity of 6883$\pm$75 \kms\ measured by van Haarlem et al. (1993) is in 
disagreement with all other values.

{\sf CGCG 160-128:} Our detection (\VHI=7920$\pm$1 \kms, \mbox{$W_{50}$}=115 \kms\ and \IHI=2.5$\pm$0.07 \Jykms) 
is consistent with our previous \nan\ detection 
(\VHI=7940$\pm$5 \kms, \mbox{$W_{50}$}=100 \kms\ and \IHI=2.2$\pm$0.3 \Jykms),
which was based on data with a 4 times higher rms of 2.8 mJy (van Driel et al. 2000).

{\sf CGCG 161-051:} We did not confirm our previous, quite tentative \nan\ detection 
(van Driel et al. 2000), with \VHI=6993: \kms, \mbox{$W_{50}$}=235: \kms\ and \IHI=1.5: \Jykms. 
Our new spectrum has a 2.4 times better rms, of 1.5 mJy.  

{\sf CGCG 161-054:} Our detection (\VHI=6756$\pm$3 \kms, \mbox{$W_{50}$}=284 \kms\ and \IHI=1.3$\pm$0.1 \Jykms) 
is consistent with our previous, tentative \nan\ detection, with 
\VHI=6760: \kms, \mbox{$W_{50}$}=335: \kms\ and \IHI=1.8: \Jykms,
which was based on data with a 4 times higher rms of 3.2 mJy (van Driel et al. 2000).

\begin{thebibliography}{}
\bibitem[]{}
Beijersbergen, M., 2003, PhD thesis, Groningen University
\bibitem[]{}
Binggeli, B., Sandage, A.,  \& Tammann, G. A., 1985, AJ, 90, 1681 (VCC)
\bibitem[]{}
Bothun, G., Aaronson, M., Schommer, R., Mould, J., Huchra, J.,  \& Sullivan, W., 1985, ApJS, 57, 423
\bibitem[]{}
Bravo-Alfaro, H., Cayatte, V., van Gorkom, J. H., \& Balkowski, C., 2000, AJ, 119, 580
\bibitem[]{}
Bravo-Alfaro, H., Cayatte, V., van Gorkom, J. H., \&  Balkowski, C.,  2001, A\&A, 379, 347
\bibitem[]{}
Byrd, G.,  \& Valtonen, M., 1990, ApJ, 350, 89
\bibitem[]{}
Chincarini, G., Giovanelli, R., \& Haynes, M. P., 1983a, \apj, 269, 13
\bibitem[]{}
Chincarini, G., Giovanelli, R., Haynes, M., \& Fontanelli, P., 1983b, \apj, 267, 511
\bibitem[]{}
Cowie, L.L.,  \& Songaila, A., 1977, Nat., 266, 501
\bibitem[]{}
de Lapparent, V., Geller, M.J.,  \& Huchra, J.P., 1986, ApJ, 302, L1
\bibitem[]{}
Dell'Antonio, I., Bothun, G.,  \& Geller, M., 1996, AJ, 112, 1759
\bibitem[]{}
Dickey, J. M., \&  Gavazzi, G. 1991, ApJ, 373, 347
\bibitem[]{}
Eder, J, Giovanelli, R., \&  Haynes, M., 1991, AJ 102, 572
\bibitem[]{}
Fontanelli, P., 1984 A\&A, 138, 85
\bibitem[]{}
Gavazzi, G., 1987, \apj, 320, 96
\bibitem[]{}
Gavazzi, G., 1989, \apj, 346, 59
\bibitem[]{}
Gavazzi, G., Carrasco, L., \& Galli, R., 1999, A\&AS, 136, 227 
\bibitem[]{}
Gavazzi, G., Boselli, A., Donati, A., Franzetti, P., \& Scodeggio, M., 2003, A\&A, 400, 451
\bibitem[]{}
Gavazzi, G., Boselli, A., van Driel, W.,  \& O'Neil, K., 2005a, A\&A, 429, 439 (Paper I)
\bibitem[]{}
Gavazzi, G., Boselli, Cortese, L., Arosio, I., Gallazzi, A., Pedotti, P., \& Carrasco, L., 
2005b (submitted to A\&A)
\bibitem[]{}
Giovanelli, R., \& Haynes, M., 1985, \apj, 292, 404
\bibitem[]{}
Giovanelli, R., (\& 23 co-authors), 2005, (AJ, in press)
\bibitem{} 
Girardi, M., Giuricin, G., Mardirossian, F., Mezzetti, \& M., Boschin, W., 1998, ApJ, 505, 74 
\bibitem[]{}
Gunn, J.E., \& Gott, J. R.III, 1972, ApJ, 176, 1
\bibitem[]{}
Haynes, M. \& Giovanelli, R., 1984, AJ, 89, 758
\bibitem[]{}
Haynes, M., Giovanelli, R., \& Chincarini, G., 1984, ARA\&A, 22, 445
\bibitem[]{}
Haynes, M., Giovanelli, R., Herter, T. et al., 1997, AJ, 113, 1197
\bibitem[]{}
Lewis, B.. Helou, G., \& Salpeter, E., 1985, ApJS, 59, 161
\bibitem[]{}
Lokas, E.L.  \& Mamon, G.A., 2003, MNRAS, 343, 401
\bibitem[]{}
Lu, N., et al., 1993, ApJS, 88, 383
\bibitem[]{}
Magri, C., 1994, AJ, 108, 896 
\bibitem{} Miller, C. J. 2004  in "Clusters of Galaxies: Probes of Cosmological Structure and Galaxy Evolution, 
from the Carnegie Observatories Centennial Symposia. Published by Cambridge University Press, 
as part of the Carnegie Observatories Astrophysics Series. 
Edited by J.S. Mulchaey, A. Dressler, and A. Oemler. 
\bibitem[]{}
Moore, B., Katz, N., Lake, G., Dressler, A.,  \& Oemler, A., Jr. 1996, Nat, 379, 613
\bibitem[]{}
Mould, J., et al., 1995, ApJS, 96, 1
\bibitem[]{}
Neumann, D. M. et al. 2001, A\&A 365, L74
\bibitem{} Nichol R.C., 2004  in "Clusters of Galaxies: Probes of Cosmological Structure and Galaxy Evolution, 
from the Carnegie Observatories Centennial Symposia. Published by Cambridge University Press, 
as part of the Carnegie Observatories Astrophysics Series. 
Edited by J.S. Mulchaey, A. Dressler, and A. Oemler, 2004, p. 24.
\bibitem[]{}
Nulsen, P.E.J., 1982, MNRAS, 198, 1007
\bibitem[]{} 
O'Neil, K., 2004, \aj, 128, 2080
\bibitem[]{}
Ramella, M., Geller, M. J., \& Huchra, J. P., 1992, ApJ, 384, 396
\bibitem[]{} 
Salzer, J., Hanson M., \& Gavazzi G., 1990, ApJ. 353, 39
\bibitem[]{} 
Scodeggio, M., \&  Gavazzi, G., 1993, ApJ, 409, 110
\bibitem[]{}
Solanes, J. et al., 2001, \apj, 548, 97.
\bibitem[]{}
Sulentic, J., \& Arp H., 1982, AJ, 88, 489
\bibitem[]{}
Sullivan III, W.T., Bothun, G.D., Bates, B., \& Schommer, R.A., 1981, AJ, 86, 919 
\bibitem[]{}
van Driel, W., Ragaigne, D., Boselli, A., Donas, J., \& Gavazzi, G., 2000, A\&AS,  144, 463
\bibitem[]{}
Vogt, N. P., Haynes, M. P., Herter, T., \& Giovanelli, R., 2004, AJ, 127, 3273
\bibitem[]{}
Williams, B., \& Kerr, F., 1981, AJ, 86, 953
\bibitem[]{}
Zwaan, M. A., et al. 2003, \aj, 125, 2842
\end{thebibliography}
\end{document}